\documentclass[prd,aps,nofootinbib,showpacs,letter,twocolumn]{revtex4-1}
\usepackage{graphicx}
\usepackage{type1cm}
\usepackage{eso-pic}
\usepackage{color}
\usepackage{amsmath}

\begin{document}

\title{Exact nonlinear inhomogeneities in $\Lambda$CDM cosmology}

\author{Nikolai Meures}
\email{nikolai.meures@port.ac.uk}

\author{Marco Bruni}
\email{marco.bruni@port.ac.uk}

\affiliation{Institute of Cosmology and Gravitation, University of Portsmouth, Portsmouth PO1 3FX, UK}

\date{\today}

\begin{abstract}
At a time when galaxy surveys and other observations are reaching unprecedented sky coverage and  precision it seems timely  to investigate  the effects of general relativistic nonlinear dynamics on the growth of structures and on observations. Analytic inhomogeneous cosmological models are an indispensable way of investigating and understanding these effects in a simplified context.

 In this paper, we develop  exact inhomogeneous solutions of general relativity  with pressureless matter (dust, describing cold dark matter) and cosmological constant $\Lambda$, which can be used to model an arbitrary initial matter distribution along one line of sight. In particular, we consider the second class Szekeres models with $\Lambda$ and split their dynamics  into a flat $\Lambda$CDM   background and exact nonlinear inhomogeneities, obtaining several new results.  One single metric function $Z$ describes the deviation from the background. We show that $F$, the time dependent part of $Z$, satisfies the familiar linear differential equation for  $\delta$, the first-order density perturbation of dust, with the usual growing and  decaying modes.  In the limit of small perturbations, $\delta \approx F$  as expected, and the growth of inhomogeneities links up exactly with standard perturbation theory. In particular, we exhibit an exact conserved curvature variable, necessary for the existence of the growing mode, which is the nonlinear extension of  the first-order curvature perturbation. We provide  analytic expressions for the exact nonlinear $\delta$ and the growth factor in our models. For the case of over-densities  we find that, depending on the initial conditions,  the growing mode may or may not lead to a pancake singularity, analogous to a Zel'dovich pancake. This  is in contrast with  the $\Lambda=0$ pure Einstein-de-Sitter background where, at any given point in comoving (Lagrangian) coordinates pancakes will always occur. Analyzing the covariant variables associated with the space-time, we derive the associated dynamical system, which we are able to decouple and reduce to two differential equations, one for $\Omega_\Lambda$ representing the background dynamics and one for $\delta$ describing the dynamics of the inhomogeneities. Our models are Petrov type D, which we show by explicitly deriving the only nonzero Weyl scalar $\Psi_2$, which does not depend on $\Lambda$. Since this is the only Weyl contribution to the geodesic deviation equation, $\Lambda$ can only contribute to lensing through its contribution to the background expansion.
 \end{abstract}
\maketitle

\section{Introduction}
From the very beginning of modern cosmology (see Einstein in \cite{LorEinMin52} and the account in \cite{Pee80}), the guiding idea behind the construction of models has been what later became known as the Cosmological Principle (cf.\  \cite{Wei72,Pee80,PetUza09,EllMaaMac11} and Refs.\ therein): the {\it assumption} that the Universe is, at any given time, homogeneous and isotropic on large scales is translated mathematically into a Robertson-Walker metric, i.e. a metric that is assumed to represent a space average and it is therefore {\it  exactly} homogeneous and isotropic. In addition, the nontrivial hypothesis is made that this metric should be a solution of Einstein's equations, thereby giving rise to a  Friedmann-Lemaitre-Robertson-Walker (FLRW) universe model. In other words, general relativity (GR) is assumed to be the correct theory describing gravitational interactions between galaxies and clusters of galaxies. While it follows logically that a Universe that is statistically homogeneous and isotropic  should be described on average by the Robertson-Walker metric, it is not at all obvious that this metric should satisfy Einstein's equations: given the nonlinear nature of the latter, averaging the equations is not the same as considering the equations satisfied by the average (see \cite{Ras06,Ras11,Buc08,Lar09} for  this ``averaging problem''). 

The Universe is however inhomogeneous. Even if it becomes homogeneous above a certain scale (see e.g.\ \cite{Sarkar:2009iga}), the size of this ``fair sample"  \cite{Pee80} depends on how we measure it and what we mean by ``statistical homogeneity"  (cf.\ \cite{Tur97} and \cite{Labini:2010aj} for alternative views). In any case,  having assumed the Cosmological Principle, the growth of inhomogeneities and their effects  are typically modeled with perturbation theory about a ``background'' FLRW model. Within this framework, the formation of nonlinear structures at smaller scales  is treated assuming a Newtonian approximation, with N-body simulations \cite{Springel:2005nw}. Most observations are interpreted assuming this Friedmannian framework; in particular,  distances are computed assuming a FLRW distance-redshift relation, i.e. completely neglecting inhomogeneities (cf.\ \cite{Okamura:2009zf} for an attempt to include them).

In the last three decades, the combination of cosmic microwave background (CMB) radiation, large scale structure (LSS) and supernova type Ia observations has provided support for a flat FLRW universe model currently undergoing an accelerated phase under the action of a dark energy component \cite{Bla10,AmeTsu10}. Consequently, the flat $\Lambda$CDM model \cite{Pee84,EfsSutMad90} has emerged as the standard concordance model of cosmology \cite{Spe03,Teg03}. GR is assumed to be the correct theory of gravity, with cold dark matter (CDM) and a cosmological constant $\Lambda$ dominating the dynamics of the late Universe. CDM is responsible for structure formation, with $\Lambda$ making up the balance to have a spatially flat Universe and driving what is inferred to be, within this conceptual framework, cosmic acceleration. Currently no other model can explain observations equally well while being statistically favored, simply because any alternative introduces more parameters, with not enough advantages (e.g.\ see  \cite{BalBruQue07,Dav07,Feb10,Hob10}). 

Einstein's equations were constructed to automatically satisfy the energy-momentum conservation equation, thus they must include a $\Lambda$ term for full mathematically generality  \cite{LorEinMin52}. Therefore, one can argue \cite{BiaRov10} that $\Lambda$ is a free parameter of the theory to be determined by observations, to be treated on the same footing as the coupling of gravity to matter - Newton's constant $G$. However, this view is not very popular and many alternatives to $\Lambda$ are considered in the literature  \cite{Copeland:2006wr,AmeTsu10,Sap10,Ras06,Buc08} in order to explain the effects that, within the homogeneity and isotropy assumption, are interpreted as acceleration \cite{Kom11}. 

We may divide these alternatives into two main groups, each with two subgroups. In the first, the Cosmological Principle is maintained, thus there is an acceleration but the cause is not a cosmological constant: \textit{i}) either $\Lambda$ is replaced by an unknown dark energy  \cite{Copeland:2006wr,AmeTsu10,Sap10} (or both CDM and $\Lambda$ are replaced by unified dark matter, see  \cite{Bertacca:2010ct,Piattella:2009kt,BerBruPei10,Lim:2010yk} and Refs.\ therein) or, \textit{ii}) an alternative theory of gravity is assumed \cite{Durrer:2008in, Koyama:2007rx}. The second alternative is to maintain GR and \textit{i}) either consider possible dynamical back-reaction effects of structure formation on the overall expansion, in order to construct an average model with acceleration \cite{Ras06,Buc08}, or \textit{ii}) consider inhomogeneous exact solutions of GR \cite{Kra97,Bolejko:2011jc} and study the effects of inhomogeneities on observations \cite{EllMaaMac11}. In the latter case, the idea is that what is normally interpreted as acceleration is at least partly due to observational effects of inhomogeneities. It turns out that, within very special models, one can even eliminate $\Lambda$ entirely, for instance for observers near the center of a spherically symmetric model (see \cite{Clarkson:2010uz}  and references therein). However, it is an open question whether this can be achieved in inhomogeneous models which exhibit an average homogeneity on the scales on which it seems to be observed \cite{Sarkar:2009iga}, i.e.\ models in which the cosmological principle is preserved.

Given the above arguments, it is of great interest to consider simple models where nonlinearities and inhomogeneities are fully taken into account while, at the same time, a FLRW background can be clearly identified. A very interesting class of models was introduced long ago by Lindquist and Wheeler \cite{LinWhe57} and revised by Redmount \cite{Red88}, which has then been recently reconsidered and generalized by Clifton and Ferreira \cite{CliFer09}, where pointlike masses (represented by Schwarzschild black holes) are distributed in a lattice and the overall expansion is described by the Friedmann equation. The main motivation of Clifton and Ferreira in following the Lindquist and Wheeler construction is the observation that indeed the Universe largely consists of galaxies and clusters of galaxies surrounded by vacuum. The question they address is how observations and measurements of the cosmological parameters are affected in a highly inhomogeneous universe whose overall dynamics is homogeneous and isotropic.
  However, this lattice construction is only an approximate solution to Einstein's equations, with ``no man's land" in between the matched spheres and, perhaps even more importantly, where the inhomogeneities are strongly nonlinear at all times. 

Having in mind the same type of questions addressed in \cite{CliFer09}, in this paper we adopt a milder approach  to the construction of models of nonlinear inhomogeneities in a clearly identifiable FLRW background.  Specifically, we consider exact solutions of GR where, starting from standard small perturbations of a FLRW universe, the matter distribution is continuous and can evolve to a highly nonlinear stage. In the process, the inhomogeneities can either form a distribution of large voids or over-densities, or a mixture of the two, with over-densities possibly even forming pancakes as in the Zel'dovich approximation in Newtonian cosmology. The benefit of our model is therefore two-fold: \textit{i}) we consider exact solutions of Einstein's equations, therefore avoiding any possible problem associated with approximations and matching and \textit{ii}) these exact solutions describe  nonlinear inhomogeneities growing on top of a FLRW background, with the possibility of modeling a rather arbitrary distribution of both voids and over-densities. It should be noted that there is no restriction on the inhomogeneities to have a zero average in any sense, thus the FLRW background may or may not be representative of an average; we shall not particularly address this issue here, leaving it for a future analysis \cite{MeuBru11}.

We do not necessarily expect that considering observational effects of nonlinear inhomogeneities can eliminate - entirely or in part -  the need for dark energy.  On the contrary, the study of the nonlinear growth of perturbations and the effects they may have on observations is a topic of the greatest interest in the context of $\Lambda$CDM cosmology.  Therefore we include a cosmological constant $\Lambda$ in our models. The exact solutions we are considering are a generalization with $\Lambda$ of the pure dust models of Szekeres \cite{Sze75} and are perse not new, as they were first found by Barrow and Stein-Schabes \cite{BarSte84}, who used them to prove a nonlinear version of the cosmic no-hair theorem\footnote{The cosmic no-hair conjecture states that, for all solutions of Einstein's equations with $\Lambda>0$ and matter satisfying the weak energy condition,  the future asymptotic state is stationary inside the cosmological event horizon of any future inextendible timelike curve.  Dynamically, this states that de Sitter space-time is the future attractor; see \cite{BarSte84,BruMenTav02} and references therein.}. However, what is new is the form in which we present this solution, which allows for a clear distinction between inhomogeneities and the FLRW background.  In doing this, we generalize to the $\Lambda\neq 0$ case the work by Goode and Wainwright \cite{GooWai82-1}, focusing on the particular subclass of solutions that admit a flat FLRW background. Through the split between FLRW background and inhomogeneities, we achieve many new results. In this paper, we present the solutions and their properties,  while we leave the analysis of the observational effects of inhomogeneities in these models to a second paper \cite{MeuBru11}.

In order to avoid any confusion with approximate solutions, in this paper we refer to \textit{deviations} (from the background) rather than \textit{perturbations}, even in the case of standard variables such as the dimensionless density inhomogeneity $\delta=(\rho-\bar{\rho})/\bar{\rho}$, where $\rho$ and $\bar{\rho}$ respectively indicate the density in the inhomogeneous universe and in the FLRW background. We choose the latter to be spatially flat, see  Eq.\  \ (\ref{eq:linek0}) for the metric  where,  in synchronous comoving coordinates, $Z$ is the single metric function that characterizes the deviations from this background.
The main results of our analysis are:
\begin{itemize}
\item generalizing the Goode and Wainwright \cite{GooWai82-1} formalism to the $\Lambda\neq 0$ case\footnote{In this paper we restrict our generalization to the second class of Szekeres-like metrics considered in  \cite{GooWai82-1}.}, we exhibit exact solutions for the growing and decaying modes of the metric deviation $Z$, assuming a flat $\Lambda$CDM background;
\item 
as for the $\Lambda=0$ case, the second order ordinary differential equation (ODE) for the growing and decaying modes of $Z$ is linear, and is the same equation that it is satisfied by $\delta$ in the linear regime; 
\item 
therefore, as for $\delta$ in the linear regime in a flat $\Lambda$CDM background, the growing mode in $Z$ asymptotically approaches  a constant value;
\item
given that the equation for the metric deviation $Z$ is linear, $Z$ satisfies a superposition principle even in the nonlinear regime, while $\delta$  does not;
\item
we explicitly show that the second order ODE satisfied by $Z$ admits a first integral; this results in a conserved curvature deviation, $d$, which turns out to be strictly related to the growing mode, i.e.\ $d=0$ implies a purely decaying mode;
\item
we eliminate the residual gauge freedom from the form of the metric in \cite{GooWai82-1}, so that $Z=1$ immediately corresponds to the FLRW background;
\item
we exhibit the exact analytic solution for the nonlinear $\delta$, showing the analogy to the Newtonian Zel'dovich pancake for the case of over-densities;
\item 
we discuss the formation of singularities at any given comoving point\footnote{Our synchronous comoving coordinates should be thought of as the equivalent of Lagrangian coordinates in a Newtonian description.} and find that, for the case of over-densities, pancakes may form in the future, while the past singularity is a FLRW-like Big-Bang for purely growing modes in $Z$, and spindlelike (cigar, or Kasner-like) if  a decaying mode is present in $Z$;
\item
given that the growing mode in $Z$ evolves to a constant, at any given point pancakes do not form if appropriate initial conditions are chosen at that point, unlike the $\Lambda=0$ case where pancakes are unavoidable;
\item
we explicitly work out the analogy between our formalism and that of perturbation theory, exploiting the linearity of the equation for $Z$ and the conserved curvature variable $d$; the latter plays the role of the conserved linear curvature perturbation;
\item
we explicitly give an analytic expression for the growth factor of the density deviation $\delta$, valid into the nonlinear regime;
\item
 we reconsider the dynamical system characterizing our models (cf.\ \cite{BruMatPan95-1,BruMatPan95-2,WaiEll05}) in terms of density $\rho$,  expansion $\Theta$, shear $\sigma_{ab}$  and Electric Weyl tensor $E_{ab}$ and, introducing new dimensionless variables, we decouple the system of ODEs, reducing the dynamics to 2 equations,  one for $\Omega_{\Lambda}$ describing the background and one for $\delta$ describing the inhomogeneities;
\item
we verify that our solutions are of Petrov type D \cite{BarRow89}, explicitly finding the null tetrad that makes $\Psi_{2}$ the only nonvanishing Weyl scalar, which we show does not directly depend on $\Lambda$: it follows that in our models $\Lambda$ can only affect lensing indirectly, through the coupling of the dynamics of the background to that of the inhomogeneities.
\end{itemize}

From this list, it is clear that some of our results are directly related to astrophysical cosmology, while others are more in the domain of relativistic cosmology. The reader more interested in the former can first read Secs.\ \ref{sec:EFEs}, \ref{sec:vissol} and \ref{sec:perts}, leaving the other sections for a second reading.

The plan of the paper is as follows. In Sec.\ \ref{sec:EFEs} we present a summary of our analysis of Einstein's field equations (EFEs); full details are given in Appendix \ref{sec:EFEdetails}. After choosing a flat $\Lambda$CDM background, we present explicit solutions for the growing and decaying modes of the metric deviation $Z$.

We give a physical interpretation to our models in Sec.\ \ref{sec:vissol}, looking at what kind of nonlinear density distributions are possible. In Sec.\ \ref{sec:sing} we present an analysis of the singularities: in particular, we look at when pancakes form or do not form, depending on the initial conditions. The space-time is analyzed using the Petrov classification scheme in Sec.\ \ref{sec:petrov}.

In Sec.\ \ref{sec:perts}, we analyze the analogy between our exact solutions and cosmological perturbation theory. In particular, in our model density inhomogeneities can grow highly nonlinear; however, we would like to use initial conditions as they are used in standard perturbation theory and hence we show how, in the linear regime, the two are related.

In Sec.\ \ref{sec:covariant}, we consider the dynamical system associated with a set of covariant variables. First, in Sec.\ \ref{sec:covder}, we relate the covariant variables with the metric functions. Then, in Sec.\ \ref{sec:covsys}, we show how the dynamical system for the covariant variables can be reduced to only two ODEs, and we present a phase plane analysis.

In Appendix \ref{sec:cont}, we demonstrate how the continuity equation directly implies the form of the density and density deviation in our model. In Appendix \ref{sec:EFEdetails}, we give details on how we solved the EFEs, with the dimensions of all the variables and parameters given in Appendix \ref{sec:dimensions}. Finally, in Appendix \ref{sec:symm}, we demonstrate how we can obtain axial symmetry in our model, depending on how we choose the free functions.

Throughout the paper, we choose units $c=8\pi G=1$.

\section{Solving the EFEs}\label{sec:EFEs}

\subsection{Setup}
In this paper we shall consider the second class Szekeres-type metrics \cite{Sze75}, in the form that Goode and Wainwright \cite{GooWai82-1} introduced
\begin{equation}\label{eq:line}
 ds^2=-dt^2+S^2\left[ e^{2\alpha(\textbf{x})}\left(dx^2+dy^2\right) + Z^2dz^2\right],
\end{equation}
where $S=S(t)$ and $Z=Z(\textbf{x},t)$. We will generalize their analysis, including a cosmological constant $\Lambda$ in Einstein's equations \cite{BarSte84}: 
\begin{equation}\label{eq:EFEs}
 G_{ab}=T_{ab}-\Lambda g_{ab}.
\end{equation}
The matter content, a pressureless dust component, represents CDM. Fluid elements move along geodesics and, in the synchronous coordinates in the metric (\ref{eq:line}), these geodesic flow lines are orthogonal to the cosmic time $t$ hyper-surfaces, with 4-velocity 
\begin{equation}
 u^a=[1,0,0,0].
\end{equation}
The coordinates in Eq.\ (\ref{eq:line}) are therefore also comoving and the fluid flow is irrotational. The energy-momentum tensor $T_{ab}$ only has one nonzero component, which is
\begin{equation}\label{eq:energy}
 T^{00}=\rho,
\end{equation}
where $\rho$ is the energy-density of the dust. It follows directly from the conservation equation ${T^{ab}}_{;b}=0$ (see Appendix \ref{sec:cont}) that in general for this metric
\begin{equation}\label{eq:rhoinit}
\rho=\frac{M(\textbf{x})}{S^3Z},
\end{equation}
where for now $M$ is a general function of space, which will be restricted by the EFEs later on. 

It should already be clear that $S=S(t)$ in the metric (\ref{eq:line}) acquires the role of a FLRW scale factor. In any case, a  simple interpretation of the metric (\ref{eq:line}) is immediately obtained if we consider the generalization of the Hubble expansion law \cite{Ell71, EllMaaMac11}. Consider two fundamental comoving observers moving with 4-velocity $u^{a}$ and connected, at any given time $t$, by a vector $X^{a}$ (thus orthogonal to $u^{a}$).
For the components of $X^{a}$ we find
\begin{subequations}
\begin{eqnarray}
\dot{X}^{x,y}&=& H X^{x,y},\\
\dot{X}^z&=&\left(H+\frac{\dot{Z}}{Z}\right)X^z,
\end{eqnarray}
\end{subequations}
where $H=\dot{S}/S$.
Hence we deduce that the Hubble law along the $x$- and $y$-axis is unmodified, i.e.\ the same as that of a FLRW background, whereas  along the $z$-axis it is changed by the inhomogeneities encoded in $Z$. As we shall see in Section \ref{sec:cov}, $\theta=\dot{Z}/Z$ is precisely the deviation of the expansion (scalar) from that of the background, $H$. Thus in our synchronous comoving representation, which is the relativistic analog of a Newtonian Lagrangian description, fluid elements occupy a fixed coordinate position, but the physical distance  between any pair of them along the $z$-axis  is modified by $Z$. As we shall see, this may even lead to pancakes, when $Z\rightarrow0$, in analogy of the Zel'dovich pancakes of the Newtonian-Lagrangian description.

\subsection{Summary of the calculations}
The details of our analysis of the EFEs for the metric (\ref{eq:line}) are given in Appendix \ref{sec:EFEdetails}. From this analysis, we obtain the following equation for the dimensionless scale factor\footnote{In the main part of the paper we use an over-dot to denote differentiation with respect to $t$. In the Appendix, however, we use a subscript $t$, for uniformity with the space derivatives in EFEs. The latter are completely solved with respect to space coordinates, hence only ODEs with respect to  $t$ need to be analyzed in the main part of the paper.}  $S(t)$:
\begin{equation}\label{eq:fried}
 \left(\frac{\dot{S}}{S}\right)^2=\frac{1}{3}\bar{\rho}+\frac{1}{3}\Lambda +\frac{K}{S^2},
\end{equation}
which we recognize as the Friedmann constraint equation for $\Lambda$CDM, where we have defined the homogeneous energy-density
\begin{equation}\label{eq:density}
\bar{\rho}=\frac{\bar{\rho}_0}{S^3};
\end{equation}
$\bar{\rho}_0$ and $S_0=1$ are the values of $\bar{\rho}$ and $S$ today. The curvature constant $K$ has dimensions and is either vanishing, positive or negative for a flat, closed or open universe respectively (see e.g.\ \cite{EllMaaMac11}); it is linked to the metric through the relation
\begin{equation}
e^\alpha=\frac{1}{1+\frac{1}{4}K(x^2+y^2)}.
\end{equation}
The function $Z$ in the line element (\ref{eq:line}) can be split as
\begin{equation}\label{eq:decoZ}
Z(\textbf{x},t)=F(z,t)+A(\textbf{x}),
\end{equation}
where $A$ can be written in the form\footnote{Note that this expression for $A$ will not be used in following Sections, since it contains gauge functions, which will be fixed in Sec.\ \ref{sec:fixgauge}.}
\begin{equation}\label{eq:decoA}
 A(\textbf{x})=a(z)+b(z)x+c(z)y+d(z)(x^2+y^2),
\end{equation}
and $F$ obeys the following linear homogeneous ODE
\begin{equation}\label{eq:diffFhom}
 \ddot{F}+2\frac{\dot{S}}{S}\dot{F}-\frac{\bar{\rho}}{2}F=0.
\end{equation}
This equation is well known: as it was noted in \cite{GooWai82-1}, it is indeed the equation satisfied by the first-order density perturbation in a dust (CDM with or without $\Lambda$) FLRW universe (see for instance \cite{Pee80}). Less well known is that this equation admits a first integral (see Appendix \ref{sec:F} for a derivation)
\begin{equation}\label{eq:difF}
 \frac{\dot{S}}{S}\dot{F}+\frac{\bar{\rho}}{2}F-\frac{2d}{S^2}=0,
\end{equation}
where, through the field equations, the conserved quantity turns out to be the curvature variable $d$, appearing in $A$, Eq.\ (\ref{eq:decoA}). Clearly Eq.\ (\ref{eq:diffFhom}) admits two linearly independent solutions:
\begin{equation}\label{eq:decoF}
 F(z,t)=\beta_+(z)f_+(t)+\beta_-(z)f_-(t),
\end{equation}
where $f_+(t)$ represents the so called growing mode and $f_-(t)$ the decaying mode. Alternatively, $f_-(t)$ is the solution to the homogeneous part of Eq.\ (\ref{eq:difF}) and $f_+(t)$ is the particular solution. We will explore in more detail the significance of the functions $S$, $f_+$, $f_-$ and the conserved quantity $d$ in the following sections.

\subsection{Solving for the metric functions}\label{sec:fixgauge}
We focus our attention on the $K=0$ case (i.e.\ a flat background) and therefore, the metric can be written in the form
\begin{equation}\label{eq:linek0}
 ds^2=-dt^2+S(t)^2\left[  dx^2+dy^2 + Z(\textbf{x},t)^2dz^2\right].
\end{equation}
Using Eq.\ (\ref{eq:density}), Eq.\ (\ref{eq:fried}) for the $K=0$ case gives
\begin{equation}\label{eq:Friedmann}
 \dot{S}^2=\frac{\bar{\rho}_0}{3S}+\frac{\Lambda}{3}S^2,
\end{equation}
which enables us to embed our space-time (\ref{eq:linek0}) in a FLRW  background through the scale factor $S(t)$. Therefore the metric (\ref{eq:linek0}) can be seen as describing an exact inhomogeneity, specified by $Z$, in the $\Lambda$CDM background described by $S(t)$ and parameterized by $\Omega_m=\bar{\rho}_0/(3H_0^2)$ and $\Omega_{\Lambda}=\Lambda/(3H_0^2)$, where $H_0$ is the Hubble parameter and $\Omega_m=1-\Omega_{\Lambda}$. Solving Eq.\ (\ref{eq:Friedmann}) then gives
\begin{equation}\label{eq:solS}
 S(t)=\left( \frac{1-\Omega_{\Lambda}}{\Omega_{\Lambda}}\right)^{1/3}\sinh ^{2/3} \left(\frac{3}{2}H_0\sqrt{\Omega_{\Lambda}}t\right),
\end{equation}
where we have set to zero an integration constant which would only shift the time of the initial singularity. With this, the age of the Universe today is
\begin{equation}
t_0=\frac{2H_0^{-1}}{3\sqrt{\Omega_{\Lambda}}}\rm{arcoth}\left(\Omega_{\Lambda}^{-1/2}\right);
\end{equation}
it is easy to check that inserting this expression into Eq.\ (\ref{eq:solS}) yields $S(0)=S_0=1$, as it should. Now, defining the dimensionless variable $\tau=\sqrt{\frac{3\Lambda}{4}}t$, the differential equation (\ref{eq:diffFhom}) for $F$ simplifies to
\begin{equation}\label{eq:Fdd}
 F''+\frac{4}{3} \coth(\tau)F'-\frac{2}{3} \frac{1}{\sinh^2(\tau)}F=0,
\end{equation}
where a dash denotes the derivative with respect to $\tau$. The two linearly independent solutions are
\begin{subequations}
\begin{eqnarray}
 f_-&=&\frac{\cosh(\tau)}{\sinh(\tau)},\label{eq:solF-}\\ 
f_+ &=&\frac{\cosh(\tau)}{\sinh(\tau)}\int\frac{\sinh^{2/3}(\tau)}{\cosh^2(\tau)}d\tau\label{eq:solF+}.
\end{eqnarray}
\end{subequations}

\begin{figure}[h!]
\begin{center}
$\begin{array}{c}
\includegraphics[width=0.4\textwidth]{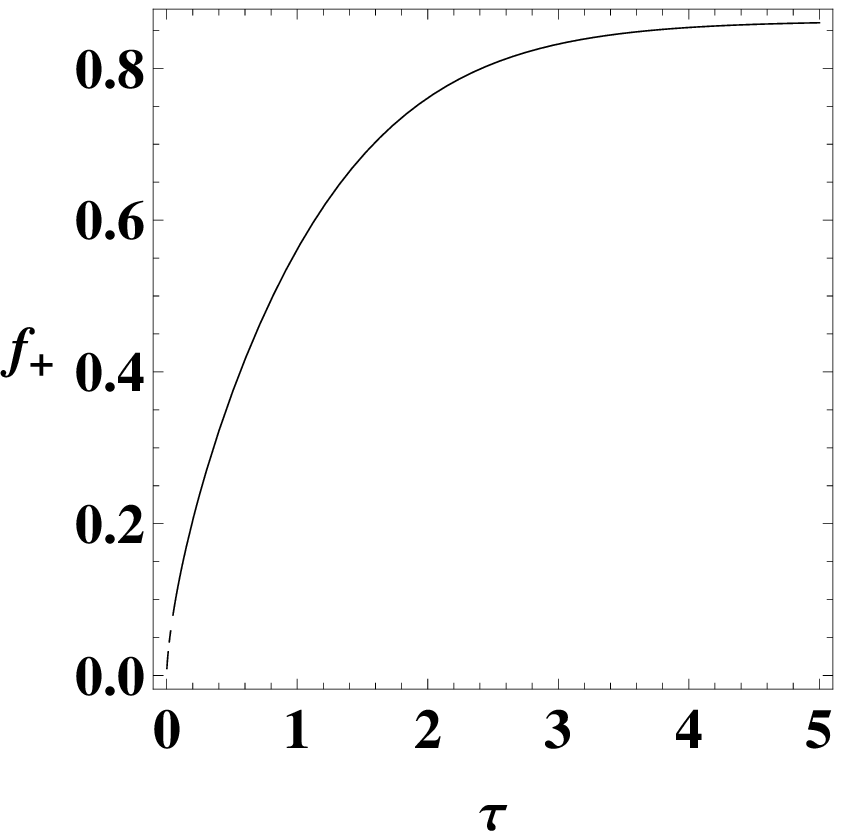}  \\
\includegraphics[width=0.4\textwidth]{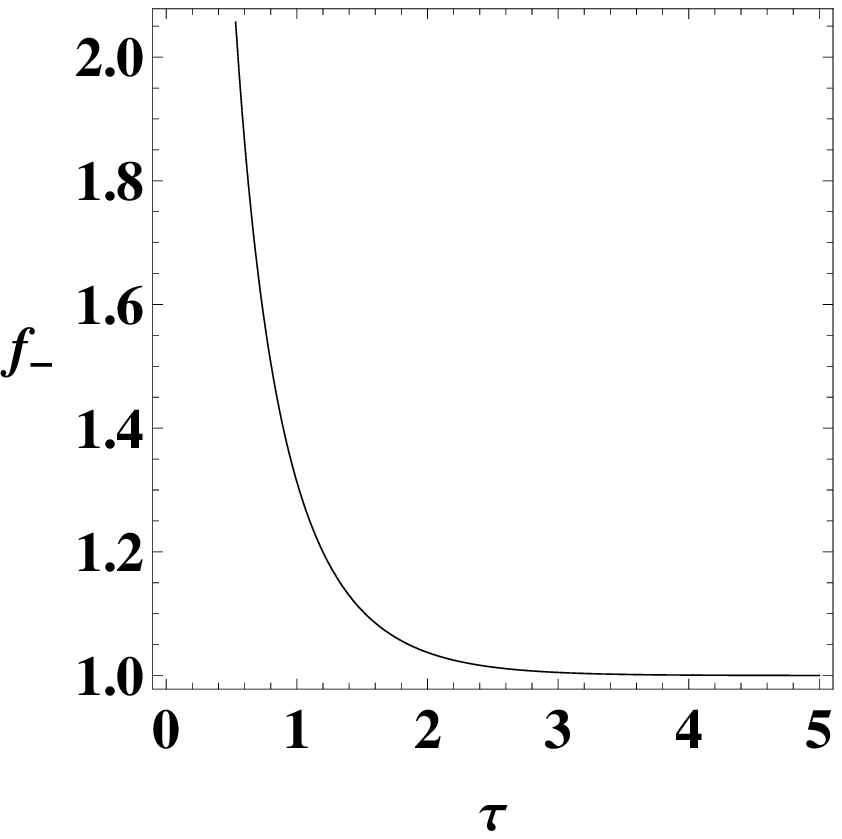}  \\
\end{array}$
\end{center}
\caption{Plots of the growing (top panel) and decaying (bottom panel) modes of the solution for $F$, as derived from Eq.\ (\ref{eq:diffFhom}). The solutions plotted here are given in Eqs.\ (\ref{eq:solF-}) and (\ref{eq:solF+}).}
\label{fig:growdecay}
\end{figure}

The function $S(t)$ in Eq.\ (\ref{eq:solS}) is the scale factor of the $\Lambda$CDM background model. The differential equation  (\ref{eq:Fdd}), obeyed by $F$, is exactly the one obeyed by the first-order density perturbation in this background (for an in depth discussion, see e.g.\  \cite{Pee80}). The two independent solutions for $F$ are shown in Fig.\ (\ref{fig:growdecay}).

It should be noted that the same formalism applies in the $\Lambda=0$ case \cite{GooWai82-1}, where  the FLRW background is an Einstein-de-Sitter model. The main difference is that the growing mode $f_+$ grows linearly with the scale factor $S$ for $\Lambda=0$, whereas in our case (see top panel in Fig.\ (\ref{fig:growdecay})), $f_+$ asymptotically approaches  a constant value.

Now, we would like to remove some residual arbitrariness in the free functions of $z$ appearing in $Z$ ($a$, $b$, $c$, $d$, $\beta_+$ and $\beta_-$). First, Eq.\ (\ref{eq:difF}) can be used to express the conserved curvature $d$ as a function of $F$ and $S$:
\begin{equation}
 d=\frac{S\dot{S}}{2}\dot{F}+\frac{\bar{\rho}_0}{4S}F.
\end{equation}

Then, one can substitute the general solution for $F$, which has been computed earlier; after some algebra one finds
\begin{equation}
 d(z)=B\beta_+(z),
\end{equation}
where we have defined
\begin{equation}\label{eq:B}
B=\frac{1}{4}\left(\bar{\rho}_0^2\Lambda \right)^{1/3}=\frac{3}{4}H_0^2\left[\Omega_{\Lambda}(1-\Omega_{\Lambda})^2\right]^{1/3},
\end{equation}
as the constant which gives dimensions of $L^{-2}$ to $d$. This can be substituted into $A$ to obtain
\begin{equation}\label{eq:Adeco}
  A(\textbf{x})=a(z)+b(z)x+c(z)y+B \beta_+(z)(x^2+y^2).
\end{equation}
Hence there are still 5 free functions of $z$ remaining in the line element ($a$, $b$, $c$, $\beta_+$ and $\beta_-$). This freedom can be further reduced by one order by transforming the $z$-coordinate. First of all, however, let us introduce two new functions $\gamma$ and $\omega$; with no loss of generality, we can write
\begin{subequations}
\begin{eqnarray}
b(z)&=& 2\gamma(z)B\beta_+(z),\\
c(z)&=& 2\omega(z)B\beta_+(z).
\end{eqnarray}
\end{subequations}
Then, substituting these in Eq.\ (\ref{eq:Adeco}) we find
\begin{equation}
 A=a+B\beta_+\left[(x+\gamma)^2\negthickspace+(y+\omega)^2-(\gamma^2\negthickspace+\omega^2)\right].
\end{equation}
Now we have to make a transformation in the $z$ coordinate and rescale $\beta_+$ and $\beta_-$. We choose
\begin{eqnarray}
 \tilde{z}&=&\negthickspace\int \negthickspace \left\{ a(z)-B\beta_+(z)\left[\gamma^2(z)+\omega^2(z)\right]\negthickspace\right\}\negthickspace dz,\\
\tilde{\beta}_+(z)&=&\frac{\beta_+(z)}{a(z)-B\beta_+(z)\left[\gamma^2(z)+\omega^2(z)\right]},\\
\tilde{\beta}_-(z)&=&\frac{\beta_-(z)}{a(z)-B\beta_+(z)\left[\gamma^2(z)+\omega^2(z)\right]}.
\end{eqnarray}
With this coordinate transformation and rescaling of $\beta_+$ and $\beta_-$, we obtain the simplification
\begin{equation}\label{eq:Adeco1}
 A(x,y,\tilde{z})=1+B\tilde{\beta}_+(\tilde{z})\left\{ \left[x+\gamma(\tilde{z})\right]^2+\left[y+\omega(\tilde{z})\right]^2\right\},
\end{equation}
and
\begin{equation}
 Z(t,x,y,\tilde{z})=\tilde{\beta}_+(\tilde{z})f_+(t)+\tilde{\beta}_-(\tilde{z})f_-(t)+\tilde{A}(\tilde{z}).
\end{equation}
With these we can now drop all the tildes and write our metric in the final form
\begin{widetext}
\begin{equation}
  ds^2=-dt^2+S(t)^2\left[dx^2+dy^2 + \left(1+\beta_+(z)f_+(t)+\beta_-(z)f_-(t)+B\beta_+(z)\left\{ \left[x+\gamma(z)\right]^2+\left[y+\omega(z)\right]^2\right\} \right)^2dz^2\right].
\end{equation}
\end{widetext}
We have thus reduced the freedom in the metric to four free functions ($\gamma$, $\omega$, $\beta_+$ and $\beta_-$). This expression also clarifies the meaning of the coordinate transformation we have just performed: in essence we have completely fixed the gauge, such that when $\beta_+$ and $\beta_-$ are equal to zero, which implies $F=0$ and $Z=1$, our metric exactly takes the form of the background FLRW space-time. Given these arguments, we will henceforth be using Eq.\ (\ref{eq:Adeco1}) as the expression for $A(x,y,z)$.

\section{Interpreting and classifying the solution}\label{sec:interp}
\subsection{Visualization of simple solutions}\label{sec:vissol}
In this section we would like to give some intuitive understanding of what kind of energy-density distributions are possible in the developed space-time. We can see from Eq.\ (\ref{eq:Mrho}) and (\ref{eq:Aeqn}) that
\begin{equation}
 A(\textbf{x})=\frac{M(\textbf{x})}{\bar{\rho}_0},
\end{equation}
therefore we can rewrite expression (\ref{eq:rhoinit}) for the density, eliminating the dependence on $M$
\begin{equation}\label{eq:rhomaster}
 \rho = \frac{M}{S^3Z}=\frac{\bar{\rho}_0A}{S^3(F+A)}.
\end{equation}
Defining as usual the dimensionless density deviation from the background density $\bar{\rho}$ (Eq.\ (\ref{eq:density}))
\begin{equation}
 \delta\equiv\frac{\rho-\bar{\rho}}{\bar{\rho}},
\end{equation}
we obtain 
\begin{equation}\label{eq:delta}
 \delta=-\frac{F}{F+A}=-\frac{F}{Z}.
\end{equation}
Using the decompositions of $F$ and $A$ derived earlier, we can write
\begin{widetext}
\begin{equation}\label{eq:decodelta}
 \delta(x,y,z,t)=-\frac{\beta_+(z)f_+(t)+\beta_-(z)f_-(t)}{1+\beta_+(z)f_+(t)+\beta_-(z)f_-(t)+B\beta_+(z)\left\{[x+\gamma(z)]^2+[y+\omega(z)]^2\right\}}.
\end{equation}
\end{widetext}
In general, the density field in (\ref{eq:decodelta}) can represent, at any given space point $\mathbf{x}$, either an over-density or an under-density, depending on the values of $\beta_+$ and $\beta_-$. However, for the case of over-densities, at any given time, there exist points where (\ref{eq:decodelta}) necessarily implies a pancake singularity. In essence, this is due to the vanishing of the function $Z$ in the denominator of (\ref{eq:delta})-(\ref{eq:decodelta}). We will discuss the existence and properties of these singularities in more detail in the next section. However, it is important to note here already that these pancake singularities are only due to the continuous description of matter in our models, i.e.\ they are shell crossing singularities (see e.g.\ \cite{Kra97}) and are analogous to Zel'dovich pancakes in Newtonian gravity.

The structure of the density distribution (\ref{eq:decodelta}) is hard to visualize in the general case $\gamma\neq0$ and/or $\omega\neq0$, which also implies the absence of Killing vectors for our space-time. Purely for illustrative purposes, we now consider the restricted case $\gamma=\omega=0$, which implies axial symmetry, see Appendix \ref{sec:symm}.

\begin{figure}[htp]
  \includegraphics[width=0.5\textwidth]{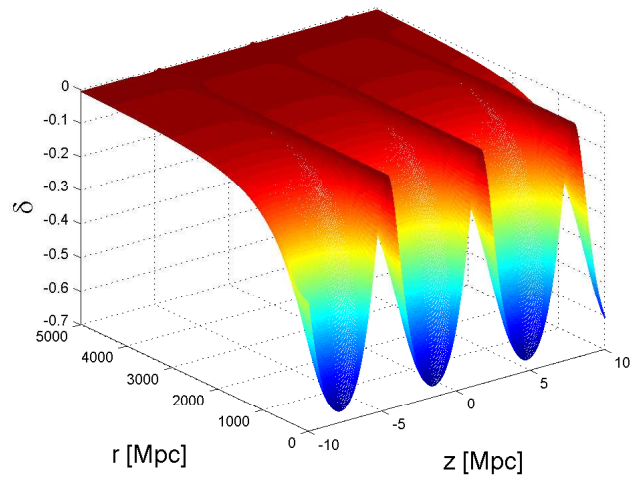}
  \caption{The $\delta$ profile of an under-density at an arbitrary time $t$ in a $\Lambda$CDM background with $\Omega_{\Lambda}=0.75$ and $H_0=72\text{km}\text{s}^{-1}\text{Mpc}^{-1}$. We assume a purely growing mode with $\beta_+\propto [1-\sin(kz)]$ for $k=1\text{Mpc}^{-1}$ and $\beta_-=\gamma=\omega=0$. In this case the space-time is axially symmetric, so $r$ is the distance from the $z$-axis, $r=\sqrt{x^2+y^2}$. All distances are comoving and given in Mpc.}\label{fig:inhomofig}
\end{figure}

As a first example of the inhomogeneous density distribution (\ref{eq:decodelta}), we consider in Fig.\ (\ref{fig:inhomofig}) a purely under-dense growing mode, $\beta_+>0$, with $\beta_-=\gamma=\omega=0$. In this case there are no pancakes and the density field is regular everywhere. The function $\beta_+$ has been chosen to take the form $\beta_+\propto [1-\sin(kz)]$ for $k=1\text{Mpc}^{-1}$ and we are plotting the density deviations at an arbitrary value of $t$. On this plot, one can see that the center of inhomogeneities runs along the $z$-axis, which is due to the condition $\gamma=\omega=0$. In general, this center can take any path around the $z$-axis, set by the two functions $\gamma(z)$ and $\omega(z)$. 

\begin{figure}[htp]
  \includegraphics[width=0.5\textwidth]{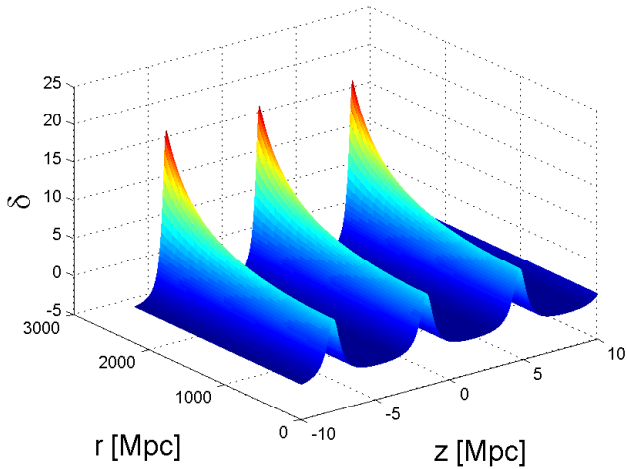}
  \caption{The $\delta$ profile for a distribution of over- and under-densities at an arbitrary time $t$ in a $\Lambda$CDM background with $\Omega_{\Lambda}=0.75$ and $H_0=72\text{~km}\text{s}^{-1}\text{Mpc}^{-1}$. We assume a purely growing mode with $\beta_+ \propto \sin(kz)$ for $k=1\text{Mpc}^{-1}$ and $\beta_-=\gamma=\omega=0$. As in Fig.\ (\ref{fig:inhomofig}), $r$ is the distance from the $z$-axis, $r=\sqrt{x^2+y^2}$. The increasing behavior of the over-densities away from the $z$-axis is due to singularities located at a certain $r_*$, beyond the boundaries of the plot. All distances are comoving and given in Mpc.}\label{fig:inhomofigoverunder}
\end{figure}

A distribution of over- and under-densities (or pure over-densities) can be simulated, if one only considers the space-time "close" to the $z$-axis, in order to avoid pancakes. An example of this is shown in Fig.\ (\ref{fig:inhomofigoverunder}). We will discuss this case in more detail in the next section.

\subsection{Classification of  singularities}\label{sec:sing}
The cosmological model we are considering contains only irrotational dust: singularities in these space-times have been studied for a long time \cite{EarLiaSac72}, see also \cite{WaiEll05}. Therefore, the types of singularities we encounter in our  model are well known, however it is essential to understand if and when they occur. 

Considering the fact that $f_+,f_-\geq0$, we can see from Eq.\ (\ref{eq:decodelta}), that if $\beta_+(z)$ and $\beta_-(z)$ assume some negative values, we have a singularity in $\delta$ if the denominator ($Z$) goes to zero. This also causes a singularity in $\rho$. This issue has been considered in great detail for the case without a cosmological constant by Goode and Wainwright \cite{GooWai82}. We define our metric functions slightly differently, such that $\beta_+(z)$ and $\beta_-(z)$ have opposite signs to the definitions in \cite{GooWai82}. The analysis of the singularities for the case $\Lambda\neq0$ gives different results than for $\Lambda=0$, which can be exploited to model inhomogeneities in a physically meaningful way.

First of all, let us introduce some formalisms. We define the variables $l_{\alpha}$ through
\begin{equation}
ds^2=-dt^2+\sum_{\alpha=1}^{3} l^2_{\alpha}\left(dx^{\alpha}\right)^2,
\end{equation}
where in our space-time we find that $l_1=l_2=S$ and $l_3=SZ$. Using these variables, one can classify singularities into three different types (see e.g.\ \cite{WaiEll05}):
\begin{enumerate}
\item a pointlike singularity when all three $l_{\alpha}\rightarrow 0$ ,
\item a cigar or spindle singularity if two $l_{\alpha}\rightarrow 0$ and the other one diverges,
\item a pancake singularity if one $l_{\alpha}\rightarrow 0$ and the other two converge to a finite value,
\end{enumerate}
as we approach the singularity. We remind the reader that we use synchronous comoving coordinates, so that in the following ``fixed space point" refers to these coordinates. 

We find that the initial singularity at $t=0$ can only be either a pointlike singularity or a cigar singularity depending only on $\beta_-$: if $\beta_-\neq 0$ we get a cigar singularity and if $\beta_-=0$ we get a pointlike singularity. In other words, if a decaying mode is present then the initial singularity is velocity dominated (see \cite{EarLiaSac72}, cf.\ also \cite{BruSop03} and references therein) and Kasner-like, while when we only have a growing mode, the initial singularity is matter dominated and effectively isotropic: in approaching the singularity, the growing mode decreases and our space-time becomes FLRW with a small perturbation. 

As we argued earlier, there can also be singularities at some time $t_*>0$, where we find divergences in the density field. We find that these singularities can only be pancake singularities. We present the singularities in Table \ref{tab:minus} for the case of $\beta_-<0$ and in Table \ref{tab:plus} for the case of $\beta_-\geq0$. 

To analyze the case $\beta_-<0$, we need a more in depth understanding of the behavior of several functions. The two time dependent functions $f_-$ and $f_+$ exhibit an asymptotic behavior for large values of $t$, as it can be seen from Fig.\ \ref{fig:growdecay}. We find $f_-\rightarrow 1$ and $f_+\rightarrow f_+^{\infty}$ as $t\rightarrow \infty$, where $f_+^{\infty}$ is a finite and positive number. This is significantly different from the $\Lambda=0$ case, where $f_+$ does not asymptote to a finite value for large $t$. Since $f_-$ and $f_+$ have an asymptotic value, it follows that for every fixed space point $F$ and hence $Z$ have a finite asymptotic value as well, see Eqs.\ (\ref{eq:decoZ}) and (\ref{eq:Adeco1}) for the definition of $Z$. We therefore introduce the new parameter $Z_{\infty}$, which is the asymptotic value of $Z$ at a given space point. We would now like to deduce that $Z$ always has a maximum value for $\beta_-<0$ and $\beta_+<0$. At early enough times, the decaying mode $f_-$ dominates over $f_+$ and so $\dot{Z}\approx\beta_-\dot{f}_-$, which is positive, since $\beta_-$ was assumed to be negative and $f_-$ is a strictly decaying function. At late enough times, we can rearrange Eq.\ (\ref{eq:difF}) to give $\dot{Z}\approx 2B\beta_+/(\dot{S}S)$, which is negative, since $B$, $S$ and $\dot{S}$ are strictly positive and we have assumed $\beta_+$ to be negative. We therefore find that $\dot{Z}$ changes sign, from positive to negative and hence must have a maximum, which we call $Z_M$. Analyzing the parameters $Z_M$ and $Z_{\infty}$ aids the distinction between different cases in Table \ref{tab:minus}.

\begin{table}[ht]
\caption{Classification of singularities occurring at some finite time $t_*>0$ for $\beta_-<0$.}
\begin{ruledtabular}
\begin{tabular}{c  c  c  c }\label{tab:minus}
$\beta_+$ & $Z_M$ & $Z_{\infty}$ & number of pancake singularities\\ \hline
$<0$&$=0$&$<0$&$1$\\
$<0$&$<0$&$<0$&$0$\\
$<0$&$>0$&$\geq 0$&$1$\\
$<0$&$>0$&$<0$&$2$\\
$\geq 0$&--&$<0$&$0$\\
$\geq 0$&--&$>0$&$1$
\end{tabular}
\end{ruledtabular}
\end{table}

In the case of $\beta_-\geq0$ we find a splitting between different cases, depending on a new parameter. Since $Z$ is positive initially ($1+\beta_-f_-$ being positive and dominating at early times), we find that if $Z_{\infty}$ is positive, $Z$ has no zeros and hence we find no pancakes. This distinction between cases turns out to depend on the value of $\beta_+$. Clearly, if $\beta_+$ is positive (and hence $Z$ is positive for all $t$), we only have under-densities and hence no singularities. If $\beta_+$ is negative, we find a critical value, which divides the cases of singularities and no singularities. This value occurs when
\begin{equation}
Z_{\infty}=1+|\beta_-|-|\beta_+|\left\{ f_+^{\infty}+ B\left[(x+\gamma)^2\negthickspace+(y+\omega)^2 \right]\right\},
\end{equation}
vanishes. For a fixed space point (and hence given values of $\beta_-$, $x$, $\gamma$, $y$ and $\omega$), we can find this critical value $|\beta_+^*|$ to be
\begin{equation}
|\beta_+^*|=\frac{|\beta_-|+1}{f_+^{\infty}+ B\left[(x+\gamma)^2+(y+\omega)^2\right]}.
\end{equation}
This parameter is used in Table \ref{tab:plus} to decide whether a given point in space will have a future singularity.

\begin{table}[ht]
\caption{Classification of singularities occurring at some finite time $t_*>0$ for $\beta_-\geq0$.}
\begin{ruledtabular}
\begin{tabular}{c  c}\label{tab:plus}
$\beta_+$ & number of pancake singularities\\ \hline
$\geq 0$& $0$\\
$<0$, $|\beta_+|<|\beta_+^*|$&$0$\\
$<0$, $|\beta_+|>|\beta_+^*|$& $1$
\end{tabular}
\end{ruledtabular}
\end{table}

The second case in Table \ref{tab:plus}, where $\beta_+<0$ and $|\beta_+|<|\beta_+^*|$, is the most interesting from a pragmatic point of view. Given a certain point in space, we can always find a value $|\beta_+^*|$, such that for $|\beta_+|<|\beta_+^*|$ there will be no future singularity at this point. This is distinctly different from the $\Lambda=0$ case where this is not possible. In practice this means that if we would like to model some density distribution on a certain space region without incurring in a pancake, we need to find the maximum $|\beta_+^*|$ within this region, which will restrict the maximum over-density we can model. "Close" to the $z$-axis this restriction (for $\beta_-=\gamma=\omega$) turns out to be fairly weak. For instance, using initial conditions at recombination, we can start with initial values even greater than the measured power spectrum amplitude \cite{Kom11} and have no future pancakes in a finite region around the $z$-axis.

\subsection{Petrov classification}\label{sec:petrov}
The Petrov classification is used to distinguish different types of space-time metrics by analyzing algebraic properties of the Weyl tensor (for a discussion of the Weyl scalars and the Petrov classification, see \cite{Cha92,Ste03}). A main point to be noted is that these properties are purely geometrical and unrelated to the theory of gravity considered. However, an understanding of these properties helps the physical interpretation, especially in those cases where the space-time can be seen as a nonlinear perturbation of some background. Such physical interpretation was given for instance by Szekeres in \cite{Sze65}, and is based on the so called Weyl scalars. These are defined by
\begin{subequations}
\begin{eqnarray}
\Psi_0&=&C_{abcd} k^a m^b k^c m^d,\\
\Psi_1&=&C_{abcd} k^a m^b k^c l^d,\\
\Psi_2&=&-C_{abcd} k^a m^b l^c \bar{m}^d,\\
\Psi_3&=&C_{abcd} l^a \bar{m}^b l^c k^d,\\
\Psi_4&=&C_{abcd} l^a \bar{m}^b l^c \bar{m}^d,
\end{eqnarray}
\end{subequations}
where $C_{abcd}$ is the Weyl tensor and $k, l, m$ and $\bar{m}$ is a complex null tetrad. These five complex scalars represent, in four dimensions, the ten components of the Weyl tensor. In essence, the Petrov classification of a certain space-time involves finding the complex null tetrad such that the number of Weyl scalars reduce to a minimal set. If one can find a tetrad such that the only non-vanishing Weyl scalar is $\Psi_2$, then the space-time is said to be Petrov type D. Well known examples are Schwarzschild and Kerr. A space-time containing gravitational waves necessarily has nonzero $\Psi_0$ and $\Psi_4$, e.g.\ a perturbed Kerr \cite{Teu74,SteWal74, Cha92}. For this reason, an analysis of the Weyl scalars can be used in extracting gravitational waves in numerical relativity, see e.g.\ \cite{Bee05,NerBeeBru05,Ner06}. On the other hand, $\Psi_0$ and $\Psi_4$ can be nonzero in a space-time with no gravitational waves, e.g.\ as in the case of a stationary rotating star \cite{Ber05}. Clearly, a Petrov type D space-time does not contain any gravitational radiation. 

The general result that Szekeres space-times are Petrov type D is well known, see e.g. \cite{BarRow89,Ste03}, therefore there are no gravitational waves. Naively, without the knowledge of the Petrov type and its meaning,  this is counter-intuitive: since we have time dependent matter inhomogeneities, one would expect gravitational radiation to be present\footnote{Another way to conclude that no gravitational radiation is present in Szekeres space-times is that the the magnetic Weyl tensor vanishes \cite{BarRow89}, see  Section \ref{sec:cov}.}. Since our metric (\ref{eq:linek0}) has the Szekeres form, it must also be Petrov type D. We now want to show this explicitly, especially to analyze the FLRW limit of our model. For a derivation of the complex null tetrad, see Section \ref{sec:covder}. Using this basis, we obtain that the only nonzero Weyl scalar is\footnote{$Z_{xx}$ represents the second derivative of $Z$ with respect to the $x$ coordinate, see Appendix \ref{sec:EFEdetails}.}
\begin{equation}
 \Psi_2=-\frac{1}{6}\left( \frac{\dot{S}}{S}\frac{\dot{Z}}{Z}+\frac{\ddot{Z}}{Z}+\frac{Z_{xx}}{S^2Z}\right).
\end{equation}
This expression is derived from the metric alone and so contains only geometric information. In particular this shows that our space-time has a single independent Weyl component. Using the EFEs, we can now relate this expression to the matter content and we find
\begin{equation}\label{eq:weyl2}
 \Psi_2=\frac{M}{6S^3Z}-\frac{\bar{\rho}_0}{6S^3}=\frac{1}{6}\bar{\rho}\delta.
\end{equation}
In the case of under-densities, it follows from Eq.\ (\ref{eq:decodelta}) that this expression goes to zero for large $x^2+y^2$ and large $t$. In the case of over-densities, at any given space point, $\Psi_2$ will diverge when a pancake forms, except in the second case in Table \ref{tab:plus}, where there is no pancake and $\Psi_2\rightarrow 0$ for large $t$. However, in approaching the pancake at $t_*$ it turns out that
\begin{equation}
\frac{\Psi_2}{\Theta^2}\simeq -\frac{1}{6}\frac{\bar{\rho}_0}{S^3}\frac{FZ}{\dot{Z}^2};
\end{equation}
see Eq.\ (\ref{eq:expsc}) below for a definition of $\Theta$. At $t_*$ all quantities in this expression have a finite value, while $Z=0$; therefore this dimensionless measure of the Weyl curvature vanishes at the pancake. A space-time with only a nonzero $\Psi_2$ is Petrov type D and a space-time with all Weyl scalars identically zero is type O, i.e.\ conformally flat. This means that we have a type D space-time in general. For all cases without a pancake, the space-time tends to a type O and FLRW space-time for large values of $x^2+y^2$ or large values of $t$. We notice from Eq.\ (\ref{eq:weyl2}) that $\Psi_2$ does not explicitly contain $\Lambda$. Since $\Psi_2$ is the only Weyl contribution to the geodesic deviation equation \cite{NerBeeBru05}, this shows that there is no direct contribution to lensing from the cosmological constant through the Weyl curvature\footnote{The issue of a direct contribution from $\Lambda$ to gravitational lensing has been the subject of recent investigations, see e.g.\ \cite{RinIsh07} and references therein.}.

\section{Relation to Perturbation Theory}\label{sec:perts}
Cosmological perturbation theory concerns itself with the dynamics of small deviations from a homogeneous FLRW background and the corresponding approximate treatment of Einstein's equations. In the covariant approach to perturbation theory \cite{EllBru89}, the variable $\Delta$ is introduced to analyze the behavior of density perturbations, see \cite{BruDunEll92,TsaChaMaa08}. This gauge-invariant variable reduces to the density perturbation $\delta$ in the comoving gauge and to the corresponding gauge-invariant variable $\Delta$ derived, in Fourier space and within the metric perturbation approach, by Bardeen\footnote{In Bardeen notation this variable is $\epsilon_m$.} \cite{Bar80} and Kodama-Sasaki \cite{KodSas84}. It can also be used in perturbative studies of the behavior of inhomogeneities and anisotropies in the neighborhood of isotropic singularities for more general space-times such as Bianchi models, see e.g.\ \cite{Dun04,AnaBru06}.

The second order differential equation governing the evolution of $\Delta$ for pressureless dust is
\begin{equation}\label{eq:Ddd}
\ddot{\Delta}+2\frac{\dot{S}}{S}\dot{\Delta}-\frac{1}{2}\bar{\rho}\Delta=0.
\end{equation}
The same equation is satisfied by $\delta$ in Newtonian theory \cite{Pee80}. In general, the dynamical content of the second order equation (\ref{eq:Ddd}) can be reexpressed by a system of two first-order equations, coupling $\Delta$ to either $C$ or $\mathcal{Z}$, due to the constraint
\begin{equation}
C=-4\dot{S}S\mathcal{Z}+2S^2\rho\Delta,
\end{equation}
where $\mathcal{Z}$ and $C$ represent the spatial variation of the expansion scalar $\Theta$ (\ref{eq:expsc}) and the 3-Ricci scalar respectively \cite{BruDunEll92}. Since we are considering irrotational dust, the system for $\Delta$ and $C$ takes the form
\begin{eqnarray}\label{eq:Dsys}
\left\{ \begin{array}{ll}
   \frac{\dot{S}}{S}\dot{\Delta}+\frac{1}{2}\bar{\rho}&\Delta=\frac{C}{S^2},\\
   &\dot{C}=0,   \end{array} \right.
\end{eqnarray}
therefore, clearly, $C$ is a quantity representing a conserved curvature perturbation, with dimensions $L^{-2}$.

For the metric developed in this paper, we find the background expansion to be given by $S$. The other time dependent function $F$ represents the deviation from homogeneity. The second order differential equation which describes its evolution is derived in Appendix \ref{sec:EFEdetails} and given in Eq.\ (\ref{eq:diffFhom}). We now notice that this is exactly the same equation as the one satisfied by $\Delta$ in Eq.\ (\ref{eq:Ddd}). The same tactic of reducing the second order equation to a set of two first-order equations can then be employed for this variable and we find
\begin{eqnarray}\label{eq:Fsys}
\left\{ \begin{array}{ll}
  \frac{\dot{S}}{S}\dot{F}+\frac{1}{2}\bar{\rho}&F=\frac{2B\beta_+}{S^2},\\
  &\left(2B\beta_+\right)\dot{}=0,  \end{array} \right.
\end{eqnarray}
where $B$ is the constant defined in Eq.\ (\ref{eq:B}). From a dynamical system perspective, it is interesting to note that the first equations in the systems (\ref{eq:Dsys}) and (\ref{eq:Fsys}) are, respectively, first integrals of Eq.\ (\ref{eq:Ddd}) and Eq.\ (\ref{eq:diffFhom}), with $C$ and $2B\beta_+$ the corresponding conserved quantities; see Appendix \ref{sec:F} for the explicit integration.

It is striking that the differential equations for $\Delta$ and $F$ take the same form. We can make the analogy even more apparent by considering the limit in which $\delta$ is small. From Eq.\ (\ref{eq:delta}) we can solve for $F$ in general
\begin{equation}
F=-\frac{A\delta}{\delta+1},
\end{equation}
so that for small values of $\delta$, $F\approx -A\delta$. This can be substituted into the differential equation for $F$ to obtain
\begin{eqnarray}
\left\{ \begin{array}{ll}
  \frac{\dot{S}}{S}\dot{\delta}+\frac{1}{2}\bar{\rho}&\delta=\frac{Q}{S^2},\\
  &\dot{Q}=0, \end{array} \right.
\end{eqnarray}
i.e.\ system (\ref{eq:Dsys}), where we have defined $Q=-\frac{2B\beta_+}{A}$. In the limit of small $\delta$ we have thus retrieved the differential equations governing the growth of density perturbations in cosmological perturbation theory, as one would expect. However, the strength of our model is that Eqs.\ (\ref{eq:diffFhom}) or (\ref{eq:Fsys}) can be used to evolve $F$ into the nonlinear regime, with Eq.\ (\ref{eq:delta}) giving the corresponding $\delta$. Moreover, we have not only found the analogy of the growth of perturbations to cosmological perturbation theory, but we have also found a conserved quantity into the nonlinear regime in our system - $2B\beta_+$. The systems of differential equations admit a decaying solution only for a zero conserved quantity, $C$  and $2B\beta_+$  respectively for the linear and exact nonlinear regimes. This shows that the growing mode solution for $F$ or $\delta$ corresponds to a particular solution to the respective equations, generated by a nonzero conserved curvature inhomogeneity, either the exact 2$B\beta_+$ or the perturbation $C$, respectively.

Finally, using (\ref{eq:solF+}) and (\ref{eq:delta}) and neglecting the decaying mode, we can write the exact nonlinear growth factor for the density inhomogeneity in our model. Defining $D_+=\delta/\delta_i$, we find
\begin{equation}
D_+=\frac{z_if_+\left\{1-\delta_i^0z_iB\left[(x+\gamma)^2\negthickspace+(y+\omega)^2 \right]\right\}}{1-\delta_i^0z_i\left\{ f_++B\left[(x+\gamma)^2+(y+\omega)^2 \right]\right\}},
\end{equation}
which, for $\omega=\gamma=x=y=0$, simplifies to
\begin{equation}
D_+=\frac{\delta}{\delta_i^0}=\frac{z_if_+}{1-\delta_i^0 z_i f_+},
\end{equation}
where $z_i$ is the redshift of the initial condition $\delta_i$ and $B$ is defined in Eq.\ (\ref{eq:B}). We have substituted the dependence on $\beta_+$ for $-\delta_i^0z_i$, where $\delta_i^0$ is the initial density perturbation along the $z$-axis. In our model we are only free to choose the distribution of $\delta$ along the $z$-axis, with the distribution of the density along the $x$- and $y$-axis then being given by the metric. This is easily understood by considering the fact that all free functions in the metric are only function of the z-coordinate. The only function in the metric containing $x$ and $y$ is $A$ with its dependence being fixed (see Eq.\ (\ref{eq:Adeco1})).

\section{Covariant variables}\label{sec:covariant}\label{sec:cov}
\subsection{Deriving the variables from the metric} \label{sec:covder}
We now consider the covariant fluid flow description of our space-times \cite{Ell71,EllBru89,EllEls99,WaiEll05,TsaChaMaa08}. For the metric (\ref{eq:linek0}) the magnetic part of the Weyl tensor $H_{ab}$ is known to be zero \cite{BarRow89} and our dust flow is irrotational, $\omega_{ab}=0$, therefore the only variables we need to consider in this section are the expansion scalar, the shear tensor and the electric part of the Weyl tensor. They are defined as, respectively,
\begin{subequations}
\begin{eqnarray}
 \Theta&=&u^a_{;a},\\
\sigma_{ab}&=&u_{(a;b)}-\frac{1}{3}\Theta h_{ab}+\dot{u}_{(a}u_{b)},\\
E_{ab}&=&C_{acbd}u^cu^d,
\end{eqnarray}
\end{subequations}
where $u_a$ is the fluid 4 velocity
\begin{equation}
 u_a=[-1,0,0,0],
\end{equation}
and we define
\begin{eqnarray}
 \dot{u}_a&=&u_{a;b}u^b,\\
h_{ab}&=&g_{ab}+u_au_b
\end{eqnarray}
and $C_{abcd}$ is the Weyl tensor. 

 From the above definition and using the metric (\ref{eq:linek0}), we find
\begin{equation}\label{eq:expsc}
 \Theta=3\frac{\dot{S}}{S}+\frac{\dot{Z}}{Z}.
\end{equation}
To analyze the shear and Weyl tensors, we need some definitions. As expected \cite{BarRow89}, from the metric we find that $H_{ab}=\omega_{ab}=0$ and hence we are considering models in the ``silent universes" class \cite{MatPanSae94}, i.e.\ cosmological models where there is no communication between fluid elements. Since the equations evolving $\rho$, $\Theta$, $\sigma_{ab}$ and $E_{ab}$ are ODEs, each fluid element evolves independently\footnote{Obviously, initial conditions have to satisfy spatial constraints, but here we are only concerned with the local time evolution.}. For an analysis of the dynamical systems for these covariant variables in ``silent models", see \cite{BruMatPan95-1} and \cite{BruMatPan95-2}, for the case with and without $\Lambda$, respectively (see also \cite{WaiEll05}).

Exact solutions in the ``silent universes" class include Bianchi I and Szekeres models, as proved in \cite{BarRow89}. The existence of more general ``silent" exact solutions of Petrov type I is unlikely for $\Lambda=0$ (see \cite{WylBer06} and references therein), but few exist for $\Lambda\neq0$ \cite{BerWyl05}. However, the ``silent" approximation with $H_{ab}=\omega_{ab}=0$ holds true for first-order scalar perturbations \cite{BruDunEll92} and at second order outside the horizon \cite{MatPanSae94-1}. It also corresponds, in the covariant description, to the so called long wavelength approximation in gravitational collapse (see \cite{DerLan95} and references therein), with $H_{ab}$ only becoming non-negligible in between different Kasner phases \cite{BruSop03}.

 In ``silent universes", according to \cite{BarRow89}, $\sigma_{ab}$ and $E_{ab}$ have a common eigenframe. We can thus expand them as
\begin{equation}
 E_{ab}=\sum_{\alpha=1}^{3}E_{\alpha}e_{\alpha a}e_{\alpha b}, \quad \sigma_{ab}=\sum_{\alpha=1}^{3}\sigma_{\alpha}e_{\alpha a}e_{\alpha b}.
\end{equation}
The orthonormal tetrad $e_{\alpha a}$ can be found to be $e_{1a}=S\delta_a^1$, $e_{2a}=S\delta_a^2$ and $e_{3a}=SZ\delta_a^3$. From this tetrad, we can now find the complex null tetrad required to compute the Weyl scalars in Section \ref{sec:petrov}. The procedure outlined in \cite{Ste04} has been followed and the complex null tetrad we find is
\begin{subequations}
\begin{eqnarray}
m_a&=&(0,\frac{S}{\sqrt{2}},-i\frac{S}{\sqrt{2}},0),\\
l_a&=&(-\frac{1}{\sqrt{2}},0,0,-\frac{SZ}{\sqrt{2}}),\\
k_a&=&(-\frac{1}{\sqrt{2}},0,0,\frac{SZ}{\sqrt{2}}),
\end{eqnarray}
\end{subequations}
and $\bar{m_a}$ is the complex conjugate of $m_a$. Returning to the shear and electric part of the Weyl tensor, they are trace free and so we write
\begin{equation}
\sum_{\alpha=1}^3E_{\alpha}= \sum_{\alpha=1}^3\sigma_{\alpha}=0.
\end{equation}
We can then define $ \sigma_+=\frac{1}{2}(\sigma_1+\sigma_2)$, $ \sigma_-=\frac{1}{2\sqrt{3}}(\sigma_1-\sigma_2)$ and $ E_+=\frac{1}{2}(E_1+E_2)$, $E_-=\frac{1}{2\sqrt{3}}(E_1-E_2)$ as a convenient set of dynamical variables, which determine the dynamics of the shear tensor and the electric part of the Weyl tensor completely. Using these definitions, we find
\begin{equation}\label{eq:sigpl}
 \sigma_+=-\frac{1}{3}\frac{\dot{Z}}{Z}, \quad \sigma_-=0,
\end{equation}
and
\begin{equation}
 E_+=\frac{\bar{\rho}_0}{6S^3}-\frac{M}{6S^3Z}=-\frac{1}{6}\bar{\rho}\delta,\quad E_-=0,
\end{equation}
where in the last equation for $E_+$ we also used the EFEs. Comparing with Eq.\ (\ref{eq:weyl2}) we see that $\Psi_2=-E_+$.

\subsection{Phase plane analysis}\label{sec:covsys}
The complete set of ODEs governing the dynamics of our models is given by
\begin{subequations}
\begin{eqnarray}
 \dot{\sigma}_+&=&-\frac{2}{3}\Theta\sigma_++\sigma_+^2-E_+,\label{eq:evosigmapl}\\
\dot{E}_+&=&-\Theta E_+-3\sigma_+E_+-\frac{1}{2}\rho\sigma_+,\\
\dot{\rho}&=&-\Theta\rho,\label{eq:consveq}\\
\dot{\Theta}&=&-\frac{1}{3}\Theta^2-6\sigma_+^2-\frac{1}{2}\rho+\Lambda.
\end{eqnarray}\label{eq:syscov}
\end{subequations}
This system of ODEs is the subset, for $\sigma_-=E_-=0$, of that considered in \cite{BruMatPan95-2,WaiEll05} for ``silent models" with $\Lambda$.

As in the previous sections, our aim is to decouple the dynamics of the background from that of the inhomogeneities. The problem is that the variables $\rho$ and $\Theta$ incorporate both a background and an inhomogeneous part. On the other hand, the shear and the electric Weyl tensor vanish in an FLRW space-time; for this reason they are first-order gauge-invariant variables \cite{EllBru89,Goo89,BruDunEll92} and here represent exact inhomogeneities. For $\Theta$ and $\rho$ we can write
\begin{eqnarray}
\Theta&=&\bar{\Theta}+\theta,\label{eq:thetadeco}\\
\rho&=&\bar{\rho}(1+\delta)\label{eq:rhodeco},
\end{eqnarray}
where, using Eq.\ (\ref{eq:expsc}), we define 
\begin{equation}
\bar{\Theta}=3\frac{\dot{S}}{S}=3H,
\end{equation}
and
\begin{equation}
\theta=\frac{\dot{Z}}{Z},
\end{equation}
and from comparing with Eq.\ (\ref{eq:sigpl}), we find $\theta=-3\sigma_+$.

Given expressions for $\Theta$ and $\sigma_+$ from the line element, one can use Eqs.\ (\ref{eq:syscov}) to find the evolution equations for $S$ and $Z$. Also, in finding $E_+$ there is an integration constant, which corresponds to $\rho_0$, yet here one finds this function to be space dependent in general. However, we can find another expression for $E_+$ from the differential equation for $\sigma_+$. We find
\begin{equation}
 E_+=\frac{2}{3}\frac{\dot{S}}{S}\frac{\dot{Z}}{Z}+\frac{1}{3}\frac{\ddot{Z}}{Z}.
\end{equation}

\begin{figure}[h!]
\begin{center}
$\begin{array}{c}
\includegraphics[width=0.5\textwidth]{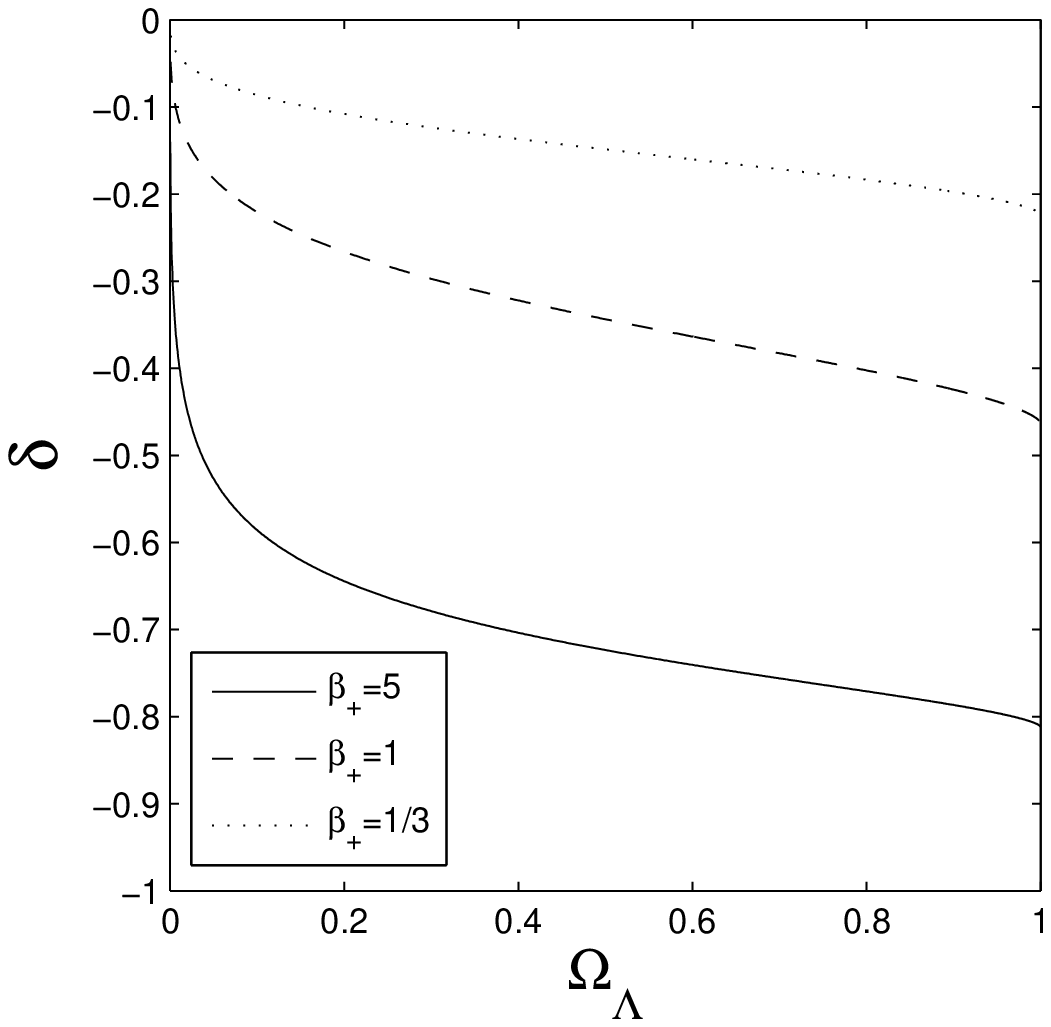}  \\
\includegraphics[width=0.5\textwidth]{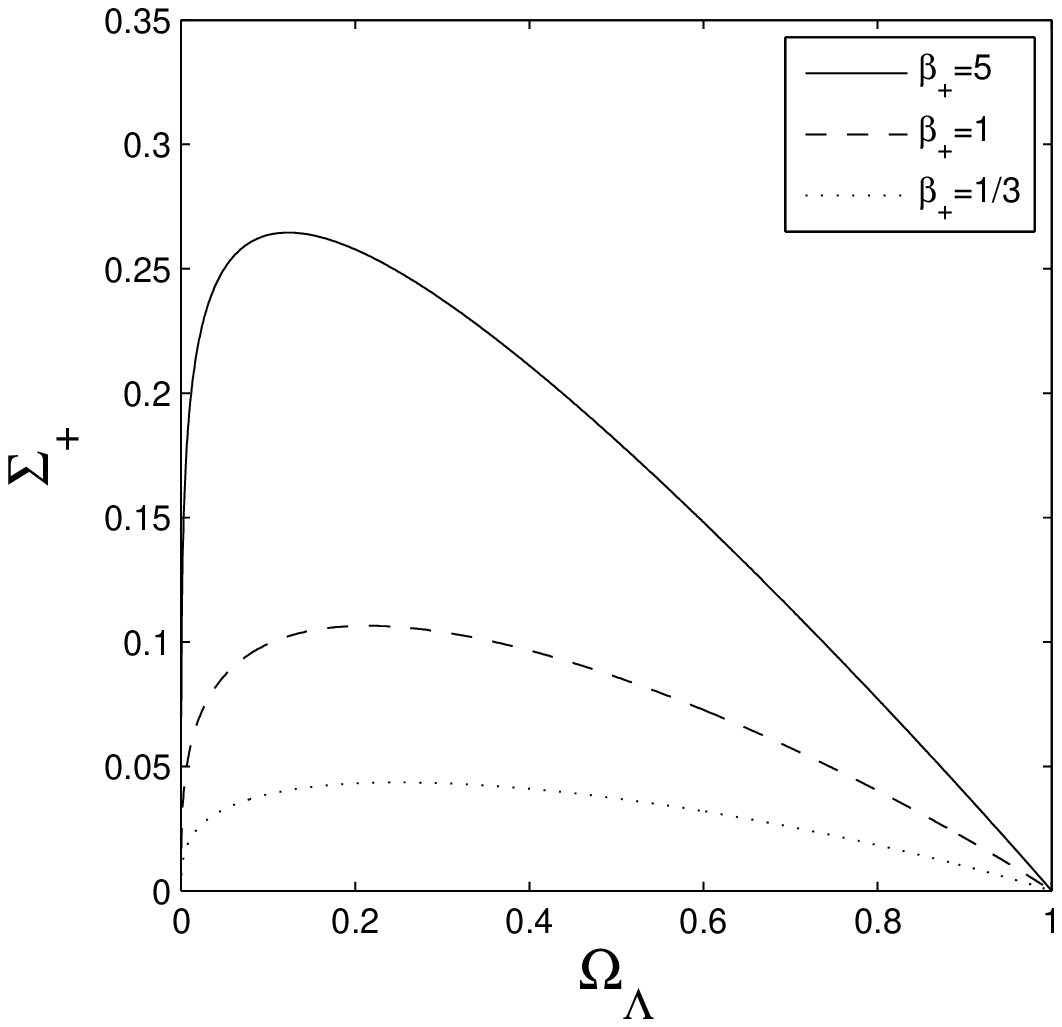}  \\
\end{array}$
\end{center}
\caption{Phase planes for the inhomogeneity variables $\delta$ and $\Sigma_+$ versus the background variable $\Omega_{\Lambda}$, for the case of under-densities, $\beta_+>0$. In each plot we only consider the growing mode and, for clarity, we plot a single trajectory for each value of $\beta_+$.}
\label{fig:phaseplane}
\end{figure}

The aim here is to describe the full dynamics of our models with as few variables and differential equations as possible. We will now need to split the covariant variables into background and inhomogeneities. To analyze the system, we start from the Friedmann equations for a flat universe; this will eventually reduce our system by one order. We start from
\begin{subequations}
\begin{eqnarray}
 \dot{H}+H^2+\frac{1}{6}\bar{\rho}-\frac{1}{3}\Lambda=0 \label{eq:Fried1},\\
H^2=\frac{1}{3}\bar{\rho}+\frac{1}{3}\Lambda.\label{eq:Fried2}
\end{eqnarray}
\end{subequations}
Combining the two equations, one finds the continuity equation for the background (see also Appendix \ref{sec:cont})
\begin{equation}
 \dot{\bar{\rho}}=-3H\bar{\rho}.
\end{equation}
Changing variables to
\begin{subequations}
\begin{eqnarray}
 \Omega_{\Lambda}&=&\frac{\Lambda}{3H^2},\\
\Omega_{m}&=&\frac{\bar{\rho}}{3H^2},
\end{eqnarray}
\end{subequations}
we can rewrite Eq.\ (\ref{eq:Fried2}) as
\begin{equation}
 1=\Omega_m+\Omega_{\Lambda}.
\end{equation}
Then, using Eq.\ (\ref{eq:Fried1}), we find
\begin{equation}\label{eq:omegald}
 \Omega_{\Lambda}'=3\Omega_{\Lambda}\left(1-\Omega_{\Lambda}\right),
\end{equation}
where $()'$ has been defined as $()'=\frac{1}{H}\dot{()}$. Eq.\ (\ref{eq:omegald}) encodes the dynamics of the background variables. From the definitions (\ref{eq:thetadeco}) and (\ref{eq:rhodeco}) and continuity equation (\ref{eq:consveq}), we obtain
\begin{equation}
 \delta ' =-\frac{\theta}{H}(1+\delta).
\end{equation}
Reexpressing the evolution equation for $\sigma_+$ (\ref{eq:evosigmapl}) in the new time and background variables we obtain
\begin{equation}
 \sigma_+'=-2\sigma_++\frac{3}{H}\sigma_+^2+\frac{H}{2}(1-\Omega_{\Lambda})\delta.
\end{equation}
We now define the new variable
\begin{equation}
 \Sigma_+=\frac{\sigma_+}{H}=-\frac{\theta}{\bar{\Theta}},
\end{equation}
which represents an expansion normalized velocity deviation. Note here that this variable $\Sigma_+$ is different to the one used in reference \cite{WaiEll05}, where the normalization of $\sigma_+$ is given by $\Theta$. We have chosen this normalization as it leads to a separation of background and perturbation variables. In addition, $H$ being a monotonic function describing the expansion of the background, the time derivative $()'=\frac{1}{H}\dot{()}$ is uniquely defined across turn-around, when $\theta=-\bar{\Theta}$. Now, consider the system
\begin{eqnarray}
\left\{ \begin{array}{lll}
 \Omega_{\Lambda}'&=&3\Omega_{\Lambda}\left(1-\Omega_{\Lambda}\right),\\
\delta'&=&3\Sigma_+(1+\delta),\\
\Sigma_+'&=&\frac{1}{2}\Sigma_+(3\Omega_{\Lambda}+1)-3\Sigma_+^2-\frac{1}{2}(1-\Omega_{\Lambda})\delta.     \end{array} \right.
\end{eqnarray}
We can further reduce the order of this dynamical system using the conserved quantity $d=\beta_+B$: using Eq.\ (\ref{eq:difF}), we can find the constraint
\begin{equation}\label{eq:constrSig}
\Sigma_+=\frac{1}{2}\left( \Omega_{\Lambda}-1\right)\delta -\frac{\beta_+}{2A}(\delta+1)\left( \Omega_{\Lambda}^{1/2}-\Omega_{\Lambda}^{3/2}\right)^{2/3}\negthickspace,
\end{equation}
where $\beta_+$ and $A$ have been defined in Eqs.\ (\ref{eq:decoF}) and (\ref{eq:decoZ}) respectively. Using this constraint, we can decouple the two differential equations for $\delta$ and $\Sigma_+$. Then, our dynamical system takes the final form
\begin{widetext}
\begin{eqnarray}
\left\{ \begin{array}{ll}
 \Omega_{\Lambda}'&=3\Omega_{\Lambda}\left(1-\Omega_{\Lambda}\right),\\
\delta'&=-\frac{3}{2}\left[\left( 1-\Omega_{\Lambda}\right)\delta+ \frac{\beta_+}{A}(\delta+1)\left( \Omega_{\Lambda}^{1/2}-\Omega_{\Lambda}^{3/2}\right)^{2/3}\right](1+\delta).    \end{array} \right.
\end{eqnarray}
\end{widetext}
Therefore, we have reduced the system of four coupled differential equations to a system of two differential equations, where the background evolution has decoupled from the deviations. The reduction in order has been due to the assumption that $K=0$ and the result that the curvature deviation is constant in our space-time, $\dot{d}=0$. The decoupling has been made possible by choosing a suitable set of variables to work with. The system is now fully characterized by the differential equations for $\Omega_{\Lambda}$ and $\delta$. $\Sigma_+$ can then easily be found using the constraint equation (\ref{eq:constrSig}). The dynamics of the variables can be shown in compressed form by doing a phase plane analysis. Plots of this analysis are shown in Fig.\ (\ref{fig:phaseplane}) for the simple case of under-densities. In the plots we have suppressed the decaying mode of $F$ for clarity.

In both the figures, the initial conditions were set very close to the $\Omega_{\Lambda}=\delta=\Sigma_+=0$ point and again for clarity, we have plotted a single trajectory for each value of $\beta_+$. This point is associated with Einstein-de-Sitter (EdS) space. The different evolutions are due to different values of $\beta_+$ and all trajectories in both figures tend to de Sitter (dS) space. In the first plot, dS space is represented by all values of $\delta$ along the $\Omega_{\Lambda}=1$ line, i.e.\ each point on this line is a fixed point\footnote{Notice that when $\delta$ is a constant, the total density $\rho$ scales with the background $\bar{\rho}\sim S^{-3}$ and therefore tends to zero at late times, cf.\ \cite{BruMenTav02} for a general second order analysis of the asymptotic evolution of perturbations in $\Lambda$CDM and the cosmic no-hair theorem that follows.}. In the second plot dS space is just represented by the $\Omega_{\Lambda}=1$, $\Sigma_+=0$ point. Therefore we see that dS is always the late time attractor for all solutions, consistently with the analysis in \cite{BarSte84}.

\section{Conclusion}

The cosmological constant problem has lead to the investigation of alternatives to the concordance $\Lambda$CDM model, considering other forms of dark energy or theories of gravity alternative to GR. However, the consequences of Einstein's equations and their nonlinearity have not yet been fully explored in cosmology. We typically use a homogeneous isotropic FLRW background plus perturbations, treating nonlinear structure formation with the Newtonian approximation.  Most observations are interpreted {\it assuming}  this Friedmannian framework, even completely neglecting inhomogeneities, as we normally do in determining distances on the base of a FLRW distance-redshift relation, cf.\ \cite{Okamura:2009zf}. 

So far there is good observational support for a  $\Lambda$CDM cosmology build in this way,  when compared with alternative theories of gravity under the same Friedmannian assumptions (see e.g.\  \cite{Daniel:2010ky,Daniel:2010yt,Bean:2010zq}), i.e. using a post-Friedmannian approach (see  \cite{PhysRevD.66.103507,Bertschinger:2006aw,Kasai:2007fn,Ferreira:2010sz} and references therein). At a time when galaxy surveys and other observations are reaching unprecedented sky coverage and  precision it seems however timely to investigate, in parallel with this linear cosmology approach to  alternative gravity theories,  the effects of  nonlinear general relativistic dynamics on the growth of structures and on observations. Analytic inhomogeneous cosmological models are indispensable to analyze and understand  these effects in a simplified context. 

Assuming GR, in this paper we have found and  analyzed  exact solutions of Einstein's equations describing an  inhomogeneous  universe with pressureless dust (representing CDM) and a cosmological constant $\Lambda$. 
Our models are of  the Szekeres class {II} type, generalized to the $\Lambda\not=0$ case; these solutions   were previously obtained in the original Szekeres form of the metric  by Barrow and Stein-Schabes \cite{BarSte84}.  We obtained our models using the metric form (\ref{eq:line}) previously used for the $\Lambda=0$ case by Goode and Wainwright \cite{GooWai82-1}. The great advantage of this form of the metric is that it allows to split the dynamics of the model into a part that describes a $\Lambda$CDM FLRW background, which we have taken to be flat, and a part describing an inhomogeneous deviation from this background, represented by  a single function $Z$ in the line element (\ref{eq:line}). The coordinates in the metric  (\ref{eq:line}) are synchronous and comoving, the relativistic analog of Newtonian Lagrangian coordinates,  thus $Z$ represents the extra change - with respect to the general Hubble expansion -  of the physical distance between pairs of fluid elements along the $z$-axis. Einstein's equations dictate that $Z=A(\textbf{x})+F(t,z)$, where the dependence of $Z$ ($A$) on the $x, y$ coordinates is fixed, while the dependence on $z$ is arbitrary. The density deviation from the background has the simple form $\delta=-F/Z$. 
This can be then used to model an arbitrary initial matter distribution along one line of sight. The coordinates in the metric  (\ref{eq:line}) are synchronous and comoving, the relativistic analog of Newtonian Lagrangian coordinates,  thus $Z$ represents the extra change - with respect to the general Hubble expansion -  of the physical distance between pairs of fluid elements along the $z$-axis.

 A very useful property of the $Z$ function is that its time dependent part, $F$, satisfies the same linear  second order ordinary differential equation (\ref{eq:diffFhom}) that $\delta$ satisfies at first perturbative order in Newtonian theory \cite{Pee80}. This same equation is satisfied by  a $\Delta$ variable in gauge-invariant perturbation theory \cite{Bar80,KodSas84,EllBru89,BruDunEll92,TsaChaMaa08}, where $\Delta$ reduces to 
$\delta$  in any comoving synchronous gauge (see \cite{Bar80, KodSas84}). This linearity property of $Z$ implies that it satisfies a superposition principle, i.e.\ we can build an arbitrary initial matter distribution along the $z$-axis, e.g.\  adding up Fourier modes, and define a spatial average. The $Z$ inhomogeneity does not need to have zero average along $z$, thus in this sense 
the FLRW background may or may not be representative of an average.  This could be used to study a simple example of averaging leading the a redefinition of the background; we leave  this and other issues for a future analysis \cite{MeuBru11}. For instance, also worth of study  is that  the linear superpositions of modes in $Z$ leads to mode coupling in the nonlinear $\delta=-F/Z$.

The two linearly independent solutions for $F$ are the well known growing and decaying modes for $\delta$ in perturbation theory \cite{Pee80}. We have shown that a crucial quantity is a conserved curvature inhomogeneity, which is a first integral of the dynamics: this generates the growing mode,  in complete analogy with perturbation theory. Assuming a vanishing decaying mode, we have given an exact formula for the nonlinear growth factor in our model.

We have also studied the local formation of singularities in our model, finding in particular that, at any given comoving point (i.e.\ for a fixed fluid element), for $\delta >0$ pancake may form, similar to the Zel'dovich pancakes of Newtonian theory. Unlike for the $\Lambda=0$ case, pancakes are not unavoidable, given that the growing mode in $Z$ for $\Lambda \not =0$ tends to a constant value.  Therefore, at any given point in our synchronous comoving coordinates, initial conditions may be found such that the pancake never forms.

The models studied here belong to the ``silent'' class \cite{BruMatPan95-1} with $\Lambda$ studied in \cite{BruMatPan95-2} (cf.\ \cite{WaiEll05} and \cite{BruSop03}). A patch in our model not evolving to a pancake unavoidably asymptotically approaches de Sitter \cite{BruMatPan95-1} (cf.\ \cite{BarSte84}), satisfying the cosmic no-hair theorem. With respect to  \cite{BruMatPan95-2, WaiEll05} we have greatly simplified the dynamical system analysis, introducing new variables that lead to a decoupling of the system, so that two variables only are needed: $\Omega_{\Lambda}$ describing the background and $\delta$ (or a shear variable) describing the inhomogeneity.

 It is known \cite{BarRow89}  that the Szekeres metric is of Petrov type D: since this is an algebraic  property of the geometry (the metric), it is unaffected by the particular form of the field equations. We have computed explicitly $\Psi_{2}$, the only non-vanishing Weyl scalar (in a specific null tetrad) for Petrov type D space-times,  showing that it does not depend on $\Lambda$. The issue of the contribution of $\Lambda$  to lensing has been a subject of debate recently, see \cite{RinIsh07} and references therein.  $\Psi_{2}$ is the only Weyl contribution to the geodesic deviation equation and $\Lambda$ contributes to the FLRW background part of the metric but not to  the inhomogeneous part of the Ricci tensor. It then follows that $\Lambda$ can only contribute to lensing indirectly, through its effect on the background expansion. 
 
 Finally, it is worth noticing that in the setting we use, of a FLRW background plus exact inhomogeneous deviations, the question arises of the gauge-invariance of these deviations, and the relation that exist to the perturbative gauge-invariance (or variance). Clearly, we have used a $\delta$ that is defined in our synchronous comoving  coordinates: while $\rho$ itself is a scalar field on the inhomogeneous space-time and as such is invariant, defining $\delta$ requires a map between the background and the inhomogeneous space-time. As it was shown in \cite{Bruni:1999et} (see also \cite{Sop04}), it turns out that scalars that vanish in the background, and that are therefore gauge-invariant at first-order \cite{EllBru89,Bruni:1996im,Sonego:1997np}, are also gauge-invariants of the exact theory\footnote{An important subtlety is that in general a field that vanishes in the background, and that is thus gauge-invariant at first-order, is not gauge-invariant at higher order, although is gauge-invariant for the exact theory: perturbations ``live in the background", the exact deviations are fields on the inhomogeneous space-time, see \cite{Bruni:1999et,Sop04} and \cite{Bruni:1996im,Sonego:1997np,Sop04}.}. The Weyl scalar $\Psi_{2}$ mentioned above is therefore an exact gauge-invariant description of the deviation between the background FLRW and the exact inhomogeneous space-time.

Having developed and analyzed our model, we would like to use it in a future work \cite{MeuBru11} to study more in the detail the growth of nonlinear density inhomogeneities and light propagation through them, to see how observations may be affected. In particular, we would like to investigate whether the distance-redshift relation is the same in an inhomogeneous space-time as in FLRW or whether non-negligible corrections need to be made. In this respect, the aim will be  to produce an analysis of this problem based on exact solutions with a continuous inhomogeneous distribution of matter; such a study will be  complementary to the work by Clifton and Ferreira \cite{CliFer09}, who used an approximate model where all matter is concentrated in a pointlike lattice distribution. Our models are certainly well suited for this task, as we have shown that nonlinear structures can be modeled along one line of sight without having to resort to any approximations. Clearly, our model can also be used to study the growth of nonlinear structures. Indeed, at least at an initial time, we can specify an arbitrary matter distribution along one direction. Simulating the growth of large over-densities can result in pancake singularities in the density field after a certain time. This is not an issue though, if one tries to simulate realistic density profiles, as we find that - in a typical $\Lambda$CDM cosmology - this problem only occurs when the perturbations during the epoch of recombination are set to be much larger than $10^{-4}$. Therefore simulating structure formation and performing light tracing along one line of sight using realistic initial conditions, based on a typical perturbation power spectrum,  is well within the scope of our model.

\appendix

\section{The continuity equation}\label{sec:cont}
In this Appendix we look at the continuity equation, at first with no assumptions on the theory of gravity and the field equations, but assuming the line element (\ref{eq:line}). First of all, let us derive the general form of $\rho$. We must solve
\begin{equation}\label{eq:conti}
\dot{\rho}=-\Theta \rho,
\end{equation}
where an expression for $\Theta$ in terms of metric functions is given in Eq.\ (\ref{eq:expsc}). Using this expression, we modify this equation to find
\begin{equation}
\frac{\dot{\rho}}{\rho}=-\frac{\left(S^3Z\right)\dot{}}{S^3Z}.
\end{equation}
Hence we can find the general solution for the density:
\begin{equation}\label{eq:rhoapp}
\rho=\frac{M(\textbf{x})}{S^3Z},
\end{equation}
which is exactly the form stated in Eq.\ (\ref{eq:rhoinit}). 

Given that $S=S(t)$, we now assume that $\rho$ can be written as 
\begin{equation}\label{eq:deltarho}
\rho=\bar{\rho}(1+\delta),
\end{equation}
where $\bar{\rho}=\bar{\rho}(t)$ is assumed to be the homogeneous density of a FLRW space-time with scale factor $S(t)$. Under this assumption $\bar{\rho}=\bar{\rho}_0S^{-3}$ and $\delta$ is the dimensionless density deviation from the background $\bar{\rho}$. Substituting this decomposition of $\rho$ into the continuity equation (\ref{eq:conti}) yields
\begin{equation}
\dot{\bar{\rho}}+3\frac{\dot{S}}{S}\bar{\rho}+\bar{\rho}\left(\frac{\dot{Z}}{Z}+\frac{\dot{\delta}}{1+\delta}\right)=0,
\end{equation}
where we identify $\dot{\bar{\rho}}=-3\frac{\dot{S}}{S}\bar{\rho}$ as the background continuity equation. Subtracting this off and rearranging, we obtain
\begin{equation}
-\frac{\dot{\delta}}{1+\delta}=\frac{\dot{Z}}{Z}.
\end{equation}
Integrating this equation we find
\begin{equation}\label{eq:deltagen}
\delta=\frac{(1+\delta_i)Z_i-Z}{Z},
\end{equation}
where $\delta_i$ and $Z_i$ are arbitrary initial values. 

If we now assume the EFEs, it follows that $Z(t,\mathbf{x})=F(z,t)+A(\mathbf{x})$, see Eq.\ (\ref{eq:decoZ2}) below, such that $Z_i=F_i+A$. With this, we can substitute into (\ref{eq:deltagen}) to obtain
\begin{equation}\label{eq:decodeltapp}
\delta=-\frac{F}{F+A},
\end{equation}
which is exactly the same form of $\delta$ as presented in Eq.\ (\ref{eq:delta}). Note that in obtaining (\ref{eq:decodeltapp}), we have assumed 
\begin{equation}\label{eq:incon}
\delta_i=-\frac{F_i}{F_i+A}.
\end{equation}
Knowing that $F$ has a decaying mode $f_-=f_-(t)$ and a growing mode $f_+=f_+(t)$ (see Appendix \ref{sec:EFEdetails} below), with the latter such that $f_+(0)=0$, we have chosen the relation between initial conditions (\ref{eq:incon}) such that, in the case of a purely growing mode, $\delta(0)=0$. Vice versa, assuming $\delta(0)=0$ implies Eq.\ (\ref{eq:incon}).

Finally, with the choice (\ref{eq:incon}), it follows from (\ref{eq:decodeltapp}), (\ref{eq:rhoapp}) and (\ref{eq:deltarho}) that 
\begin{equation}\label{eq:Mrho}
M=\bar{\rho}_0A,
\end{equation}
which we used in Eq.\ (\ref{eq:rhomaster}).

\section{Details on solving the EFEs}\label{sec:EFEdetails}
In this Appendix we solve EFEs (\ref{eq:EFEs}) for the line element (\ref{eq:line}) with the energy momentum tensor given in (\ref{eq:energy}). Since we are only considering irrotational dust and a cosmological constant, we can use the off-diagonal terms of $G_{ab}$ as constraints. We start by finding
\begin{eqnarray}
 G_{tx}&=&-\frac{Z_{xt}}{Z}=0,\\
 G_{ty}&=&-\frac{Z_{yt}}{Z}=0.
\end{eqnarray}
This implies that we can write
\begin{equation}\label{eq:decoZ2}
Z(\textbf{x},t)=F(z,t)+A(\textbf{x}).
\end{equation}
Furthermore we find
\begin{equation}
 G_{tz}=\frac{2\alpha_zZ_t}{Z}=0,
\end{equation}
and since we know that $Z$ will change with time we find that
\begin{equation}\label{eq:alpha}
 \alpha(\textbf{x})=\alpha(x,y).
\end{equation}
The last restriction from off-diagonal terms is given by the $G_{xy}$ term. It dictates
\begin{equation}\label{eq:restric1}
 \frac{-Z_{xy}+\alpha_yZ_x+\alpha_xZ_y}{Z}=0;
\end{equation}
we will use this constraint later. The four diagonal terms give the equations
\begin{subequations}
\begin{widetext}
\begin{eqnarray}
G_{tt}=&2\frac{S_t}{S}\frac{Z_t}{Z}+3\left(\frac{S_t}{S}\right)^2-\frac{1}{S^2e^{2\alpha}}\left( \frac{Z_{xx}}{Z}+\frac{Z_{yy}}{Z}+\alpha_{xx}+\alpha_{yy}\right)&=\rho +\Lambda\label{eq:eqn1},\\
\frac{G_{xx}}{S^2}=&2\frac{S_{tt}}{S}+\left(\frac{S_t}{S}\right)^2+3\frac{S_t}{S}\frac{Z_t}{Z}+\frac{Z_{tt}}{Z}+\frac{1}{ZS^2e^{2\alpha}}\left( \alpha_yZ_y-\alpha_xZ_x-Z_{yy}\right)&=\Lambda\label{eq:eqn2},\\
\frac{G_{yy}}{S^2}=&2\frac{S_{tt}}{S}+\left(\frac{S_t}{S}\right)^2+3\frac{S_t}{S}\frac{Z_t}{Z}+\frac{Z_{tt}}{Z}+\frac{1}{ZS^2e^{2\alpha}}\left( \alpha_xZ_x-\alpha_yZ_y-Z_{xx}\right)&=\Lambda\label{eq:eqn3},\\
\frac{G_{zz}}{S^2Z^2}=&2\frac{S_{tt}}{S}+\left(\frac{S_t}{S}\right)^2-\frac{1}{S^2e^{2\alpha}}(\alpha_{yy}+\alpha_{xx})&=\Lambda\label{eq:eqn4}.
\end{eqnarray}
\end{widetext}
\end{subequations}
From this system, one can see that by subtracting Eqs. (\ref{eq:eqn2}) and (\ref{eq:eqn3}) from each other, one obtains
\begin{equation}\label{eq:restric2}
 -2\alpha_xZ_x+2\alpha_yZ_y+Z_{xx}-Z_{yy}=0.
\end{equation}
Multiplying Eq.\ (\ref{eq:eqn4}) by $S^2S_t$ gives
\begin{equation}
 2S_{tt}SS_t+S_t^3-\frac{S_t}{e^{2\alpha}}(\alpha_{xx}+\alpha_{yy})=\Lambda S^2S_t.\label{eq:20}
\end{equation}
Together with Eq.\ (\ref{eq:alpha}), given that $S=S(t)$, this shows that the term $e^{-2\alpha}(\alpha_{xx}+\alpha_{yy})$ is a constant in space and time. Integrating this equation with respect to time then gives
\begin{equation}
 SS_t^2-\frac{S}{e^{2\alpha}}(\alpha_{xx}+\alpha_{yy})=\frac{1}{3}\Lambda S^3 +\mathcal{C},
\end{equation}
where $\mathcal{C}$ is an integration constant. Rearranging this equation one finds
\begin{equation}\label{eq:fried2}
 \left(\frac{S_t}{S}\right)^2=\frac{\mathcal{C}}{S^3}+\frac{1}{3}\Lambda +\frac{K}{S^2},
\end{equation}
which is the Friedmann constraint equation for $\Lambda$CDM. We have defined
\begin{equation}
 K\equiv-e^{-2\alpha}(\alpha_{xx}+\alpha_{yy}),
\end{equation}
the constant term found in (\ref{eq:20}). This differential equation admits the solution
\begin{equation}
 e^\alpha=\frac{1}{1+\frac{1}{4}K(x^2+y^2)}.
\end{equation}
Thus we find for $K=0$ that $e^\alpha=1$ and therefore $\alpha_x=\alpha_y=\alpha_{xx}=\alpha_{yy}=0$. From now on we restrict our attention to a flat universe.

In the Friedmann equation (\ref{eq:fried2}), we identify the term $\mathcal{C}/S^3$ with the energy-density term $\bar{\rho}/3$. This satisfies the background continuity equation (see Appendix \ref{sec:cont}) that gives 
\begin{equation}
\bar{\rho}=\frac{\bar{\rho}_0 }{S^3},
\end{equation}
and therefore we identity our constant $\mathcal{C}$ as 
\begin{equation}\label{eq:bigC}
3\mathcal{C}=\bar{\rho}_0.
\end{equation}
Since we are only considering a flat universe, we find
\begin{equation}\label{eq:ZXY}
 Z_{xy}=0,
\end{equation}
and
\begin{equation}\label{eq:ZXX}
 Z_{xx}=Z_{yy},
\end{equation}
from Eqs.\ (\ref{eq:restric1}) and (\ref{eq:restric2}) respectively. We can thus reduce the system Eqs.\ (\ref{eq:eqn1})-(\ref{eq:eqn4}) to
\begin{subequations}
\begin{eqnarray}
2\frac{S_t}{S}\frac{Z_t}{Z}+3\left(\frac{S_t}{S}\right)^2\negthickspace-2\frac{Z_{xx}}{ZS^2}=\rho&+&\Lambda\label{eq:eqn31},\\
2\frac{S_{tt}}{S}+\negthickspace\left(\frac{S_t}{S}\right)^2 \negthickspace +3\frac{S_t}{S}\frac{Z_t}{Z}+\frac{Z_{tt}}{Z}-\frac{Z_{yy}}{ZS^2}&=&\Lambda\label{eq:eqn32},\\
2\frac{S_{tt}}{S}+\left(\frac{S_t}{S}\right)^2\negthickspace&=&\Lambda\label{eq:eqn33}.
\end{eqnarray}
\end{subequations}
The combination $(\text{\ref{eq:eqn32}})+\frac{1}{2}(\text{\ref{eq:eqn33}})-\frac{1}{2}(\text{\ref{eq:eqn31}})$ gives
\begin{equation}\label{eq:2SZ}
 3\frac{S_{tt}}{S}+\frac{Z_{tt}}{Z}+2\frac{S_t}{S}\frac{Z_t}{Z}=-\frac{\rho}{2}+\Lambda.
\end{equation}
Combining Eqs.\ (\ref{eq:fried2}), (\ref{eq:bigC}) and (\ref{eq:eqn33}) yields
\begin{equation}\label{eq:difffried}
 \frac{S_{tt}}{S}+\frac{\bar{\rho}_0}{6S^3}-\frac{\Lambda}{3}=0,
\end{equation}
which is the Friedmann equation for $\Lambda$CDM. Subtracting 3 times (\ref{eq:difffried}) from (\ref{eq:2SZ}) gives
\begin{equation}\label{eq:2Z}
 \frac{Z_{tt}}{Z}+2\frac{S_t}{S}\frac{Z_t}{Z}+\frac{M}{2S^3Z}-\frac{\bar{\rho}_0}{2S^3}=0,
\end{equation}
where we have used Eq.\ (\ref{eq:rhoapp}). Substituting the decomposition of $Z$ we found in Eq.\ (\ref{eq:decoZ2}), we find
\begin{equation}\label{eq:gzrel}
 S^3F_{tt}+2S^2S_tF_t-\frac{\bar{\rho}_0}{2}F=-\frac{M}{2}+\frac{\bar{\rho}_0}{2}A.
\end{equation}
Clearly the LHS is a function of time and $z$ only and the RHS is a function of $x$, $y$ and $z$ only. Hence both sides must be equal to a function of $z$ only. Call this function $g(z)$: then $F$ satisfies the differential equation
\begin{equation}\label{eq:diffFpart}
 F_{tt}+2\frac{S_t}{S}F_t-\frac{\bar{\rho}_0}{2S^3}F=\frac{g(z)}{S^3}.
\end{equation}
This ODE has two homogeneous and one particular solution. We denote the homogeneous solution $F^{\rm h}=F^{\rm h}(z,t)$. The particular solution is easily spotted to be
\begin{equation}
 F^{\rm p}=-\frac{2g(z)}{\bar{\rho}_0}.
\end{equation}
This gives the form
\begin{equation}
 F=F^{\rm h}(z,t)-\frac{2g(z)}{\bar{\rho}_0}.
\end{equation}
Also, from Eq.\ (\ref{eq:gzrel}) we find
\begin{equation}
 A=\frac{2g(z)}{\bar{\rho}_0}+\frac{M}{\bar{\rho}_0},
\end{equation}
which implies
\begin{equation}
 Z=F^{\rm h}(z,t)+\frac{M}{\bar{\rho}_0}.
\end{equation}

This shows that the metric is completely independent of the function $g(z)$. This can be understood by looking back at Eq.\ (\ref{eq:decoZ2}). There we have decomposed the function $Z$ into two separate functions $A$ and $F$, but both those functions are functions of $z$, which means that there is always a certain arbitrariness in the choice of $F$ and $A$. We could easily add $g(z)$ to $F$ and subtract it from $A$ and still end up with the same function $Z$. Thus, we can choose here the function $g(z)$ to be equal to zero without loss of generality, which gives
\begin{equation}\label{eq:Aeqn}
 A=\frac{M}{\bar{\rho}_0}.
\end{equation}
Keeping the function $g(z)$ would not change any results but would merely clutter the equations. In other words, we only need the homogeneous part of Eq.\ (\ref{eq:diffFpart}) to completely specify our solution. From now on therefore, we will assume that $F$ satisfies the equation
\begin{equation}\label{eq:diffFhom2}
 F_{tt}+2\frac{S_t}{S}F_t-\frac{\bar{\rho}_0}{2S^3}F=0.
\end{equation}
As first pointed out in \cite{GooWai82-1}, this linear ODE is the equation satisfied by first-order density fluctuations in a perturbed dust FLRW universe.

Since this equation has two linearly independent solutions, which exhibit growing and decaying behavior, we can write $F$ as
\begin{equation}
 F(z,t)=\beta_+(z)f_+(t)+\beta_-(z)f_-(t),
\end{equation}
where $f_+(t)$ and $f_-(t)$ are the growing and decaying solutions respectively and $\beta_+(z)$ and $\beta_-(z)$ are free functions of $z$. Using this decomposition of $F$ and Eqs.\ (\ref{eq:ZXY}) and (\ref{eq:ZXX}) we can find the functional form of $A$, obtaining
\begin{equation}
 A(\textbf{x})=a(z)+b(z)x+c(z)y+d(z)(x^2+y^2).
\end{equation}
Looking at the system of differential equations, Eqs.\ (\ref{eq:eqn31})-(\ref{eq:eqn33}), we can see that we have only extracted two equations out of this system, so we should be able to get more information out of it. Subtracting Eq.\ (\ref{eq:eqn32}) from Eq.\ (\ref{eq:eqn33}) we find
\begin{equation}
 3\frac{S_t}{S}\frac{Z_t}{Z}+\frac{Z_{tt}}{Z}-\frac{Z_{xx}}{ZS^2}=0,
\end{equation}
which can be brought into a slightly different form by using the relation between $Z$ and $F$,
\begin{equation}
 3\frac{S_t}{S}F_t+F_{tt}-\frac{Z_{xx}}{S^2}=0.
\end{equation}
Using Eq.\ (\ref{eq:diffFhom2}) and noting that $Z_{xx}=2d$, we find
\begin{equation}\label{eq:difF2}
 \frac{S_t}{S}F_t+\frac{\bar{\rho}_0}{2S^3}F-\frac{2d}{S^2}=0.
\end{equation}
This equation really takes the form of a first integral equation of Eq.\ (\ref{eq:diffFhom2}), see Appendix \ref{sec:F} below. We discuss the form of this equation in more detail in Sec.\ \ref{sec:perts}, on perturbation theory.

\section{The relation between the first and second order equations for $F$}\label{sec:F}
We have found two different differential equations for $F$, one first-order (\ref{eq:difF2}) and one second order (\ref{eq:diffFhom2}). We explicitly show here that the first-order equation is the first integral of the second order equation when $K=0$. We start by modifying Eq.\ (\ref{eq:diffFhom2}) to obtain
\begin{equation}\label{eq:moddiffF}
\frac{S_t}{S}F_{tt}+2\left(\frac{S_t}{S}\right)^2F_t-\frac{S_t}{S}\frac{\bar{\rho}}{2}F=0.
\end{equation}
From Eqs.\ (\ref{eq:fried2}) and (\ref{eq:difffried}) we can find 
\begin{equation}
2\left(\frac{S_t}{S}\right)^2=\frac{\bar{\rho}}{2}+\frac{S_{tt}}{S}+\left(\frac{S_t}{S}\right)^2.
\end{equation}
Using this expression, we can rewrite Eq.\ (\ref{eq:moddiffF}) as
\begin{equation}
SS_tF_{tt}+SS_{tt}F_t+S_tS_tF_t+\frac{\bar{\rho}_0}{2S}F_t-\frac{\bar{\rho}_0}{2S^2}FS_t=0,
\end{equation}
which can easily be integrated to obtain
\begin{equation}
\frac{S_t}{S}F_t+\frac{\bar{\rho}_0}{2S^3}F-\frac{C}{S^2}=0,
\end{equation}
where $C$ is a constant in time. This equation takes the same form as Eq.\ (\ref{eq:difF2}). However the function $C$ here cannot be related to the metric, whereas, when this differential equation is derived using the EFEs, we can find that $C=d(z)$, which features in the metric itself.

\section{Dimensional Analysis}\label{sec:dimensions}
We add this Appendix to aid the reader in gaining some physical interpretations of some of the variables. In this section $L^n$ denotes the dimension of length to the power of $n$; assuming $c=8\pi G=1$, we find
\begin{equation}
[Z]=L^0,
\end{equation}
\begin{equation}
[t]=L,
\end{equation}
\begin{equation}
[  ,_t]=L^{-1},
\end{equation}
\begin{equation}
[x]=[y]=[z]=L,
\end{equation}
\begin{equation}
[S]=L^0,
\end{equation}
\begin{equation}
[\bar{\rho}_0]=L^{-2},
\end{equation}
\begin{equation}
[\Lambda]=L^{-2},
\end{equation}
\begin{equation}
[B]=L^{-2},
\end{equation}
\begin{equation}
[\tau]=L^0.
\end{equation}

\section{Symmetries}\label{sec:symm}
We would like to show here that for certain choices of the free functions of the metric, we find axial symmetry. From the line element Eq.\ (\ref{eq:linek0}), we find the metric
\begin{equation}
 g_{ab}= \left(
\begin{array}{cccc}
-1&0&0&0\\
0&S(t)^2&0&0\\
0&0&S(t)^2&0\\
0&0&0&S(t)^2Z(t,x,y,z)^2\\
\end{array}
\right).
\end{equation}
Clearly this metric is dependent on all 4 space-time variables. We can decompose $Z(t,x,y,z)$ as
\begin{equation}
 Z = 1+F(z,t)+B\beta_+\left[ (x+\gamma)^2+(y+\omega)^2\right],
\end{equation}
where $\beta_+=\beta_+(z)$, $\gamma=\gamma(z)$ and $\omega=\omega(z)$. If we choose $\gamma=\omega=0$ (choosing $\gamma=c_1$ and $\omega=c_2$ gives the same result) we find
\begin{equation}
 Z(x,y,z)=1+F(z,t)+B\beta_+(z)\left( x^2+y^2\right).
\end{equation}
By making the coordinate transformation
\begin{equation}
x=\rho \sin(\phi), \quad y=\rho \cos(\phi),
\end{equation}
we can rewrite $Z$ as
\begin{equation}
 Z(\rho, z)=1+F(z,t)+B\beta_+(z)\rho^2.
\end{equation}

In the new coordinates $(t, \rho, \phi,z)$ we find
\begin{equation}
 g_{ab}= \left(
\begin{array}{cccc}
-1&0&0&0\\
0&S(t)^2&0&0\\
0&0&S(t)^2\rho^2&0\\
0&0&0&S(t)^2Z(t,\rho,z)^2\\
\end{array}
\right).
\end{equation}
The metric does not depend on $\phi$, hence, for $\gamma=\omega=0$, the solution has an axial symmetry about the $z$-axis.

\bibliographystyle{apsrev4-1}

%

\end{document}